\documentclass[acmsmall]{acmart}  

\def\BibTeX{{\rm B\kern-.05em{\sc i\kern-.025em b}\kern-.08emT\kern-.1667em\lower.7ex\hbox{E}\kern-.125emX}}

\usepackage[redeflists]{IEEEtrantools}
\usepackage{graphicx}
\usepackage{textcomp}
\usepackage{xcolor} 

\usepackage{subcaption}

\usepackage{comment}
\usepackage{pbox}

\usepackage{array}
\usepackage{enumitem} 
\usepackage[utf8]{inputenc}
\usepackage[T1]{fontenc}
\usepackage{booktabs, multirow, bigstrut}
\usepackage{siunitx}
\usepackage[flushleft]{threeparttable} 
\usepackage{pifont} 
\usepackage[T1]{fontenc} 
\usepackage{makecell, caption} 
\usepackage{rotating, graphicx} 
\usepackage{makecell} 
\usepackage{adjustbox}
\newlength\mylength
\acmJournal{CSUR}
\acmYear{2021}
\acmDOI{10.1145/1122445.1122456}

\begin{document}

\title{A Taxonomy of Cyber Defence Strategies Against False Data Attacks in Smart Grid}
\author{Haftu Tasew Reda}
\affiliation{%
  \institution{La Trobe University}
   \city{Melbourne}
  \state{Victoria}
  \country{Australia}}
\email{h.reda@latrobe.edu.au}

\author{Adnan Anwar}
\affiliation{%
  \institution{Deakin University}
  \city{Geelong}
  \state{Victoria}
  \country{Australia}}

\author{Abdun Naser Mahmood}
\affiliation{%
 \institution{La Trobe University}
 \city{Melbourne}
 \state{Victoria}
 \country{Australia}}
 
 \author{Zahir Tari}
\affiliation{%
 \institution{RMIT University}
 \city{Melbourne}
 \state{Victoria}
 \country{Australia}}

\renewcommand{\shortauthors}{H. T. Reda et al.}
\begin{abstract}
Modern electric power grid, known as the Smart Grid, has fast transformed the isolated and centrally controlled power system to a fast and massively connected cyber-physical system that benefits from the revolutions happening in the communications and the fast adoption of Internet of Things devices. While the synergy of a vast number of cyber-physical entities has allowed the Smart Grid to be much more effective and sustainable in meeting the growing global energy challenges, it has also brought with it a large number of vulnerabilities resulting in breaches of data integrity, confidentiality and availability. False data injection (FDI) appears to be among the most critical cyberattacks and has been a focal point interest for both  research and industry. To this end, this paper presents a comprehensive review in the recent advances of the defence countermeasures of the FDI attacks in the Smart Grid infrastructure. Relevant existing literature are evaluated and compared in terms of their theoretical and practical significance to the Smart Grid cybersecurity. In conclusion, a range of technical limitations of existing false data attack detection researches are 
identified, and a number of future research directions are recommended. 
\end{abstract}



\maketitle
\section{Introduction} \label{sec:introduction}
\IEEEPARstart{E}{nergy} is the backbone of our economic growth and is a super-critical resource on which all other national critical infrastructure sectors rely upon. Significant rise in threats to critical infrastructure from nation states and malicious actors poses real challenges in the understanding of operational vulnerabilities in a Smart Grid as well as the different attack vectors that may jeopardize the stability and efficiency of the power system. The 2020 Global Risks report by the World Economic Forum \cite{12} indicates that large scale cyberattacks against critical infrastructure and networks is the top most threat and will continue to be among the most likely global threats over the next 10 years. 

\begin{figure}
	\centerline{\includegraphics[width=9.5cm]{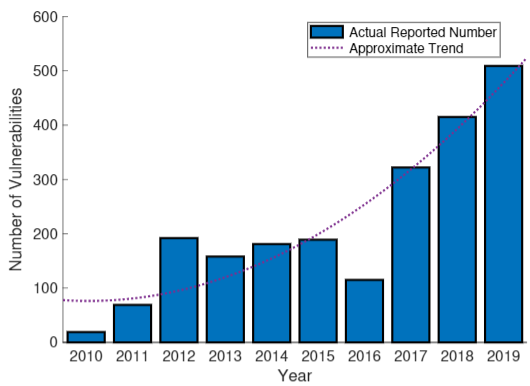}}
	\caption{Number of ICS vulnerabilities by year (reproduced from the US ICS-CERT \cite{11} and Kaspersky ICS-CERT \cite{13})}
	\label{vulEnergy}
\end{figure}

According to vulnerability reports from the US ICS-CERT \cite{11} and Kaspersky ICS-CERT \cite{13}, the energy sector has reported the largest number of vulnerabilities among all critical infrastructures. For example, Fig. \ref{vulEnergy} shows the number of vulnerabilities of various Industrial Control System (ICS) elements between 2010 and 2019 \cite{11} \cite{13}. Accordingly, 178, 110, and 283 cyberattack incidents were recorded in the energy sector out of 322, 415, and 509 ICS cyberattack incidents, respectively across the fiscal years 2017, 2018, and 2019. These cyber incidents may lead to myriads of security risks including the loss of critical data necessary for control operations, malicious modification/deletion of critical power system states. Possible consequences include incorrect customer billing information, price manipulation in the energy market, small to large scale electric power outage, and the likelihood of endangering lives by limiting power to other national critical infrastructures. 

This paper discusses various state-of-the-art false data injection (FDI) \cite{liu2011false} defence countermeasures in Smart Grid.
\subsection{Purpose and Scope of the Study}
Bad data detection (BDD) \cite{6655273} \cite{7868276} \cite{7232283} has been widely utilized in the power system control centers for the identification of cyber anomalies. Nevertheless, it has been proven that the BDDs are incapable of detecting FDI \cite{liu2011false} attacks. The extensive studies on potential FDI attacks have enabled Smart Grid operators to set up a range of defence mechanisms. The primary objective of this article is to provide a systematic literature review and insights into a taxonomy of various defence countermeasures against cyber-physical attacks in cyber-physical system. 
\subsection{Contributions}
This article has analyzed related and recent publications and reference materials in the mitigation techniques of the false data attacks across various domains of the Smart Grid infrastructure. We systematically search for older and more recent related literature, analyse the main findings covered in each literature, critically evaluate them, and compare each solution within the broader conception of the cyber-physical data integrity attacks. Specifically, major contributions of this article are summarised below. 
\begin{enumerate}
\item The paper identifies essential cybersecurity requirements of Smart Grid (Section \ref{CPSattacks}), including a theoretical analysis of stealthy FDI attacks, and requirements for the stealthy FDI attacks (Section \ref{fdiAtt1}). 
\item After a thorough review of relevant existing survey papers, this work highlights their contribution and identifies the gaps that have been addressed through this survey. Detailed comparisons have been highlighted in Table \ref{relW} and the related discussions have been presented in Section \ref{countermeasure}.   
\item The paper comprehensively analyses defence categories (Section \ref{countermeasure}) and compiles a list (Table \ref{defTable}) of methodologies including statistical, signal processing, and advanced AI-based methodologies which can be used to detect and prevent the FDI attacks.
\item In addition, this paper analyses the various countermeasure methods and provides statistical
 facts on the basis of the evaluation criteria in Section \ref{comparisonSec} and Table \ref{criteriaTable}. Furthermore, the paper discusses main research gaps in the existing papers in Section \ref{litGap}.
\item Finally, this paper provides technical recommendations for emerging advanced application areas, including Internet of Things (IoT)-based Advanced Metering Infrastructure (AMI), cognitive radio, lightweight ML for resource-constrained IoT devices, distributed attack detection in edge computing environment, and Blockchain-based defence for privacy preservation in the Smart Grid.
\end{enumerate}
\subsection{Outline of the Paper}
First, Section \ref{relWor} discusses related survey papers on defence countermeasures of the false data attacks and compares with our paper. Next, background on Smart Grid and key cyber-physical elements are discussed in Section \ref{backG}. Then, cyber-physical attacks, cybersceurity main objectives, and security requirements of Smart Grid are highlighted in Section \ref{CPSattacks}. Further, in Section \ref{fdiAtt1}, we comprehensively discuss the FDI attack, the attack vector construction methodologies, and the  main requirements for the FDI attack under the Smart Grid environment. Section \ref{countermeasure} presents the suggested taxonomy and the defence strategies against the false data attacks that are critical frameworks for the power system operator and other stakeholders. Literature search methodology, selection \& analysis of the surveyed literature, and evaluation criteria among the multitude of algorithms of selected surveyed papers are presented in Section \ref{revMethod}. Furthermore, we compare and contrast among the numerous defence strategies in Section \ref{comparisonSec}. Following a critical review of the shortcomings found in the literature in Section \ref{litGap}, our technical recommendations that can substantiate future researches in the field are provided in Section \ref{futurePros}. Finally, Section \ref{cncln} concludes this survey article.
\section{Related Survey Papers}\label{relWor}
The work by Z Guan et. al \cite{10.2015.066756} is one of the earliest works where authors present a comprehensive survey of  attack and defence of the FDI. \cite{10.2015.066756} has overviewed detection schemes and presented on the basis of centralised-and distributed-based SE techniques. Furthermore, a survey research of the data injection attacks with respect to three major cybersecurity aspects, namely FDI attack construction, impacts of the attacks, and countermeasures is studied by R Deng et. al \cite{7579185}. Unlike to previous studies, \cite{7579185} thoroughly studied the impacts of data injection attacks on the electricity market. Another line of survey research is studied in \cite{LIU201735}, which summarises related literature on different attack models, economic impact of the attack, and mitigation techniques for various Smart Grid domains including transmission, distribution, and microgrid networks. Moreover, G Liang et. al \cite{7438916} complement previous studies and discuss various FDI attack models, physical and economic impacts of the attacks, and countermeasures in Smart Grid. Research works in \cite{8744664} and \cite{8766775} also comprehensively discuss the FDI attacks from the attacker's and operator's point of view along with the consequential impacts of the attacks. 

Different from previous surveys the authors of \cite{8887286} reviewed two main classes of detection algorithms: model-based and data-driven, and have discussed the benefits and drawbacks of each technique. As compared to other review works which mostly focus on the energy management system (EMS), the authors in \cite{AOUFI2020102518} discussed FDI attacks on various entities of the online power system security. These authors review and compare studies on the FDI attacks and provide a new class of cyber-oriented countermeasure: prevention (further classified into block chain and cryptography based techniques).

Unlike to the related works, this paper presents a detailed survey of recent developments in the FDI and sets out a taxonomy of the incumbent cyberattack with respect to defence strategies across every Smart Grid domain including transmission to consumption, AGC to microgrids/DERs, substation to wide area monitoring systems. IoT, cognitive radios, and software-defined networks have recently been introduced as enablers to the Smart Grid. These communication technologies are very important to address the cybersecurity aspects of today's Smart Grid which were missed in most of the existing related works. In general, in light of research, this paper provides an in-depth survey of the latest advances of the defence measures against the cyber-physical FDI attacks within the Smart Grid infrastructure. Table \ref{relW} summarises the comparison of existing survey papers and this article.
\begin{table*}
\caption{Comparison of current survey articles and our paper}
\label{relW}
\begin{center}
\scalebox{0.80}{
\begin{tabular}{|m{1cm}m{1cm} |m{0.5cm}|m{0.5cm}|m{0.5cm}|m{0.5cm}|m{0.5cm}|m{0.5cm}|m{0.5cm}|m{0.5cm}|m{0.5cm}|} 
\cline{1-11}
& &  \multicolumn{9}{c|}{\textbf{Literature}} \\ \cline{3-11}
\multicolumn{2}{|c|}{\textbf{Comparison attributes}} & \cite{10.2015.066756} & \cite{7579185} & \cite{7438916} & \cite{LIU201735} & \cite{8744664} & \cite{8766775} & \cite{8887286} & \cite{AOUFI2020102518} & \pbox{0.8cm}{Our paper} \\ \cline{1-11}
\multicolumn{1}{|c}{\multirow{3}{*}
{\textbf{Defence based on SE type}}} &
\multicolumn{1}{|c|}{Conventional BDD}  &\ding{53}& \ddag&$\checkmark$ & \ddag&\ding{53} &\ddag &$\checkmark$ &\ddag & $\checkmark$ \\ \cline{2-11} &
\multicolumn{1}{|c|}{Detection based on dynamic SE} &\ding{53} &\ding{53} &\ding{53} &\ddag & \ding{53}&\ddag & $\checkmark$& \ddag& $\checkmark$ \\ \cline{1-11}
\multicolumn{1}{|c}{\multirow{4}{*}
{\textbf{Protection-based defence}}}&
\multicolumn{1}{|c|}{Optimal PMU placement} & \ding{53}&$\checkmark$& $\checkmark$& $\checkmark$& $\checkmark$&\ddag& \ddag&\ddag & $\checkmark$ \\ \cline{2-11} &
\multicolumn{1}{|c|}{Optimal measurement selection} &\ding{53} & $\checkmark$&\ding{53} & \ddag&\ding{53} &\ding{53} &\ding{53} &\ddag & $\checkmark$ \\ \cline{2-11} &
\multicolumn{1}{|c|}{Grid topology perturbation} &\ding{53} &\ding{53} & \ding{53}& \ddag &\ding{53} &\ding{53} &\ddag & \ddag& $\checkmark$ \\\cline{1-11}
\multicolumn{1}{|c}{\multirow{6}{*}
{\textbf{Statistical-based detection}}}&
\multicolumn{1}{|c|}{GLR test detector}  &\ddag & \ddag &\ding{53} &\ding{53} & \ding{53}&\ding{53} &\ddag  &\ding{53} & $\checkmark$ \\ \cline{2-11} &
\multicolumn{1}{|c|}{Bayesian test detector}  &\ddag &\ding{53} &\ding{53} &\ding{53} & \ding{53}& \ding{53}&\ding{53} & \ding{53}& $\checkmark$ \\ \cline{2-11} &
\multicolumn{1}{|c|}{Quickest change detector}  &\ddag &\ding{53} &\ddag &\ding{53}  & \ding{53}& \ddag& \ddag & \ding{53}& $\checkmark$ \\ \cline{2-11} &
\multicolumn{1}{|c|}{Statistical distance}  & \ding{53}&\ding{53} & \ddag& \ding{53} &\ding{53} & \ddag& \ddag &\ddag & $\checkmark$ \\ \cline{2-11} &
\multicolumn{1}{|c|}{Sparse matrix recovery} & \ding{53} &\ding{53} & \ddag&\ddag & \ding{53}& \ding{53}&\ding{53} &\ding{53} & $\checkmark$ \\ \cline{1-11} 
\multicolumn{1}{|c}{\multirow{6}{*}
{\textbf{Data-driven dection}}}&
\multicolumn{1}{|c|}{Supervised ML} & \ding{53}&\ding{53} & \ding{53}&\ding{53} &\ding{53} &\ddag & $\checkmark$&\ddag & $\checkmark$ \\ \cline{2-11}&
\multicolumn{1}{|c|}{Semi-supervised ML} &\ding{53} &\ding{53} &\ding{53} & \ding{53} &\ding{53} & \ddag & $\checkmark$ &  \ding{53}& $\checkmark$ \\ \cline{2-11} &
\multicolumn{1}{|c|}{Deep learning} & \ding{53}& \ding{53}& \ding{53}& \ding{53} & \ding{53}&\ddag &\ddag & \ddag& $\checkmark$ \\ \cline{2-11} &
\multicolumn{1}{|c|}{Reinforcement learning}  & \ding{53}& \ding{53}& \ding{53}& \ding{53} &\ding{53} & \ddag&\ddag & \ding{53}& $\checkmark$ \\ \cline{2-11} &
\multicolumn{1}{|c|}{Deep reinforcement learning} &\ding{53} &\ding{53} &\ding{53} &\ding{53} & \ding{53}&\ding{53} &\ding{53} &\ding{53} & $\checkmark$ \\ \cline{1-11} 
\multicolumn{1}{|c}{\multirow{2}{*}
{\textbf{Prevention-based defence}}}&
\multicolumn{1}{|c|}{Cryptographic-based prevention} & \ding{53} & \ding{53} &  \ding{53}& \ding{53} & \ddag & \ding{53} & \ding{53} & $\checkmark$ & $\checkmark$ \\ \cline{2-11}&
\multicolumn{1}{|c|}{Blockchain-based prevention} & \ding{53} & \ding{53} & \ding{53} & \ding{53} & \ding{53} & \ding{53} & \ding{53} & $\checkmark$ & $\checkmark$ \\ \cline{1-11}
\multicolumn{2}{|c|}{\textbf{Evaluation criteria}} &
 \ding{53} & \ding{53} & \ding{53} & \ding{53} & \ding{53} & $\ddag$ & $\ddag$ & $\ddag$ & $\checkmark$ \\ \cline{1-11}
\multicolumn{2}{|c|}{\textbf{Future directions}}  & \ding{53} & \ding{53} & $\checkmark$ & $\checkmark$ & \ding{53} & $\checkmark$ & $\checkmark$ & $\checkmark$ & $\checkmark$ \\ \cline{1-11} 
\multicolumn{2}{|c|}{\textbf{Duration of surveyed papers}} & \pbox{0.5cm}{2009 to 2013} & \pbox{0.5cm}{2009 to 2015} & \pbox{0.5cm}{2009 to 2015} & \pbox{0.5cm}{2009 to 2016} & \pbox{0.5cm}{2010 to 2017} & \pbox{0.5cm}{2010 to 2019} & \pbox{0.5cm}{2011 to 2019} & \pbox{0.5cm}{2009 to 2019} & \pbox{0.5cm}{2009 to 2020} \\ \cline{1-11}
\end{tabular} }
\begin{tablenotes}
  \item \: \: \: $\checkmark$ studied/covered,  \ddag \:partially studied, \ding{53} not studied 
  \end{tablenotes}
\end{center}
\end{table*}
\section{Background} \label{backG}
Smart Grid is the convergence of various cyber and physical components of the electrical power domain. In other words, it is an evolution of the electrical power system that is co-engineered through expertise from different fields such as OT (namely physical power devices, data acquisitions, control systems, and industrial automation), IT (namely decision making and human interfaces), advanced ICT infrastructure, and cybersceurity to be more effective and sustainable in meeting the growing global energy challenges. As compared to the traditional power grid, Smart Grid provides an end-to-end system of two-way electricity flow in which customers cannot only utilize energy but can also feed energy back into the grid. Smart Grid also supports wide variety of energy sources (including the renewables which are key to low-carbon emissions).    
According to NIST's \cite{1213} conceptual model, Smart Grid comprises of seven seven interconnected application domains: generation, transmission, distribution, customer, market, service provider, and operator. 

While applications in every Smart Grid domain are critical to the scalability, efficiency and stability of the power system operations, they also introduce vulnerabilities to the Smart Grid. The probability of a successful breach is inevitable for all cyber-physical systems directly or indirectly linked to the Smart Grid. Hence, it is crucial to scrutinize the relations between the physical model and the cyber system in order to provide a resilient cyber and communications infrastructure in the Smart Grid environment. Therefore, in this section, we briefly discuss the main cyber-physical elements of the Smart Grid.
\subsection{SCADA}
Supervisory control and data acquisition (SCADA) \cite{101} \cite{suaboot2020taxonomy} is an industrial and power system control application. Usually a SCADA consists basically of three subsystems: a data acquisition sub-system that collects measurement of the power system, a supervisory sub-system that has the ability to control remote  intelligent electronic devices (IEDs) \cite{101} by transmitting control commands (such as to close/open a circuit breaker, to change a transformer tap settings, to lower/raise generator output, and etc), and a communication sub-system that interconnects the data acquisition sub-system to the supervisory sub-system. A typical scenario in the integrated SCADA system can be described, for example, when the SCADA gathers data from diverse IEDs in a power system through various communication methods (such as IP-based wide area networks, local area networks, and software-defined networking), and then control/monitor the data using different visualisation tools.
\begin{figure*}[!ht]
\centerline{\includegraphics[width=140mm, scale=1.2]{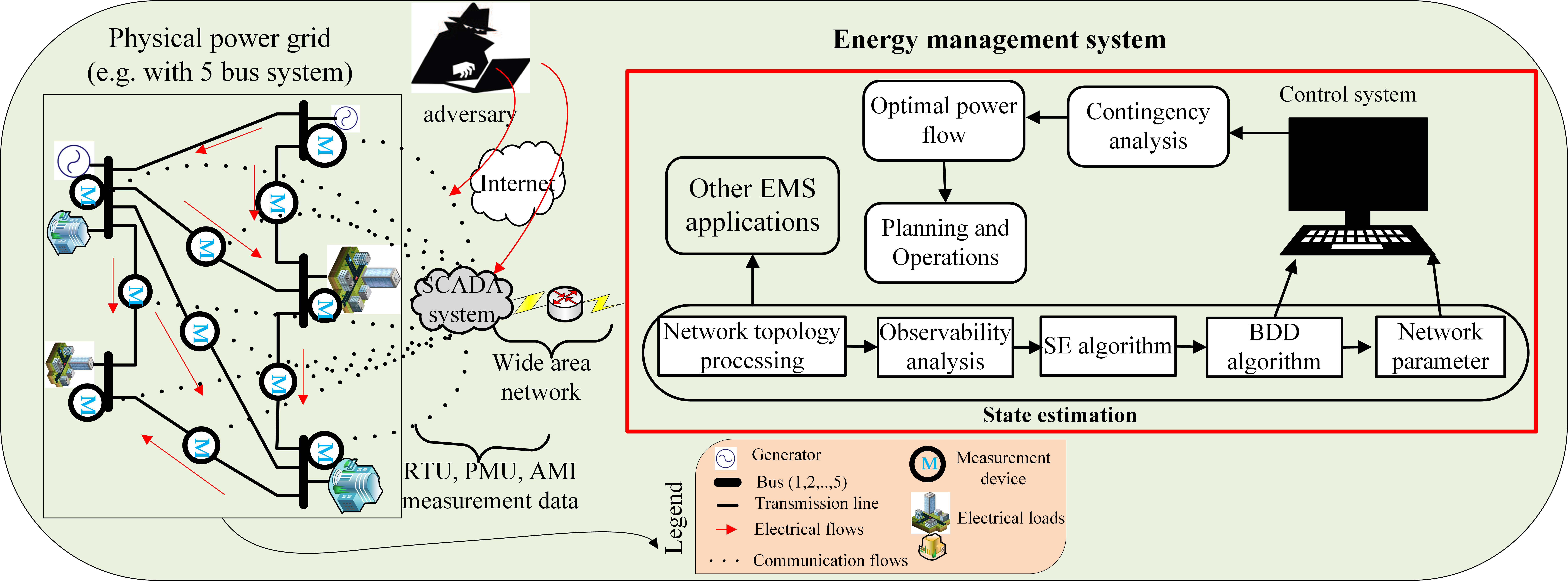}}
\caption{Typical Smart Grid with 5-bus system}
\label{motSE}
\end{figure*}
\subsection{Energy Management System}
Power system operations are regulated by system operators from the control center. Within the control center lies EMS, an automation system used to monitor, control, coordinate, and optimize energy data performance across the majority of Smart Grid infrastructure in real time. EMS depends on a SCADA system for its data monitoring
and analysis events. Distribution grids (usually from substation to consumption-side) are controlled via the distribution energy management (DEM) system. 

A typical EMS comprises the following functional elements including state estimation (SE), optimal power flow, contingency analysis (CA), alarm management system, planning and operations, automatic generation control and economic dispatch. Generally, a physical power grid can be considered as a set of buses, transmission lines, loads, generators, shunt components, and etc. Each of the buses or nodes are physically interconnected by lines or branches. Fig. \ref{motSE}\footnote{This typical Smart Grid consists of 5 buses, 11 smart power meters, communication links, and SCADA communication system. An intruder can compromise the power measurements, mislead outcome of the EMS, induce abnormality in the power system operation, and can lead to power outage.} is a typical Smart Grid architecture using IEEE 5-bus\footnote{IEEE 5 bus system is a linearized DC real power flow data which approximates real-world electric power systems using 5 buses and 17 sensors, accounting for the nodal power injections and line power flows. https://ieeexplore.ieee.org/document/5589973} system. It illustrates how the EMS and other ICT components communicate with the physical power grid.
\subsection{Smart Grid Communication Systems}
Communication systems are essential to the efficient operation of the Smart Grid. Various communication technologies are utilised across the different domains. For example, IEC 61850 in substation automation system (SAS), PMU in wide area monitoring systems (WAMS), AMIs across customer-side, and NCS
between sensors, actuators and controllers. 
\subsection{Distributed Energy Resources (DERs)}
DERs are decentralised, versatile, and modular architecture that incorporate a number of renewable sources, including solar, wind, geothermal, etc. Compared to conventional approaches in which energy is generated by centralised and big power plants, DER now allows energy production and delivery from many areas, including millions of homes and businesses. Microgrid technology is one of the enablers of Smart Grid that provides smooth collaboration between DERs offering isolation options (also known as 'islanding') or access to the conventional grid electricity.
\section{Cyber-Physical Security of Smart Grid} \label{CPSattacks}
The security issues of Smart Grid have emerged from both
physical and cyber spaces that include:
 physical security \cite{1213} (i.e. security policies with respect to staffs or personnel, physical equipment protection, and contingency analysis), 
  cybersecurity (focusing on the information security of Smart Grid pertaining to IT, OT, network and communication systems), and
  cyber-physical security (incorporating strength in all physical and cybersecurity measures against inadvertent cyber-physical incidents within an integrated Smart Grid framework). 
In this section, Smart Grid cybersecurity goals, cybersecurity requirements, and cyber-physical attacks are highlighted.
\subsection{Smart Grid Cybersecurity Goals}
Quality of service and secure power supply are the primary concern of power companies and industrial sectors. So much that the Smart Grid strives to build a much more efficient and reliable energy, cybersecurity threats can inevitably slow down its progress. Therefore, the Smart Grid needs to ensure the basic security goals such as data integrity, availability confidentiality, accountability, and etc of the various cyber-physical elements. While these security principles have been developed to govern policies on generic information security within organisations, the principles of Smart Grid cybersecurity have also been identified by NIST \cite{1213}.

\textbf{Avaiablity}:
The permanent availability and timeliness of electricity are crucial in our day to day life. Within the Smart Grid environment, availability is by far the most critical security goal for stability of the power grid. It ensures reliable access to and timely use of information. Availability can be quantified in terms of latency, the time required for data to be transmitted across the power grid. Smart Grid cybersecurity solutions should provide acceptable latency thresholds of various applications by minimising detrimental effects on the availability.

\textbf{Integrity}:
Integrity is the second yet highly critical Smart Grid security requirement. As part of the cybersecurity objectives, integrity ensures that data should not be altered without authorized access, source of data need to be verified, the time stamp 
linked with the data must be identified/validated, and quality of service is under acceptable range.

\textbf{Confidentiality}:
From the point of view of system reliability, confidentiality seems to be the least important as compared to availability and integrity. Nevertheless, with the proliferation of smart meters and AMIs across the Smart Grid implies the increasing importance of confidentiality to prevent unauthorized disclosure of information, and to preserve customer privacy or proprietary information.   

\textbf{Accountability}: 
Another security objective within the Smart Grid ecosystem is accountability, a requirement that consumers should be responsible for the actions they take. Accountability is very important, particularly when customers obtain their billing information from the utility center, they will have sufficient evidence to prove the total power load that they have used.
\subsection{Smart Grid Security Requirements}
The dynamics of the cyber-physical interaction in the Smart Grid poses extrinsic system dependencies. Further, the open inter-connectivity of Smart Gird with the Internet brings various security challenges. Therefore, Smart Grid requires stringent holistic security solutions to uphold the security objectives discussed above and to provide salient features within the Smart Grid infrastructure. First of all, the security solutions need to be robust enough to counteract against increasing security breaches that can lead to loss of data availability, loss of data integrity, loss of data confidentiality. In other words, the operation of power system should continue 24/7 regardless of cyber incident maintaining the power grid reliability (consistent to the data availability and to almost 99.9\% \cite{1213} of data integrity across the power system), and ensure consumer privacy. Second, resilient cyber-physical operations are required. According to NIST's recommendation \cite{calder2018nist}, cybersecurity in critical infrastructure such as the Smart Grid can adopt a comprehensive security framework containing five main features. These include identifying of risks or cyber incidents, providing protective mechanisms against the impact of a potential cybersecurity event, providing defence mechanisms to allow prompt discovery of security breaches, appropriate response to minimise the effect of the incident, and recovery plans to restore any systems that have been disrupted due to cyber accidents. Moreover, as attacks from cyber criminals on the power grid continue to rise in complexity and frequency, it is inevitable that various parts of the Smart Grid are vulnerable to the incumbent attacks. Therefore, it is required to provide strong attack defence across the EMS and to deploy secure communication protocols.  
\section{False Data Injection Attacks} \label{fdiAtt1}
FDI attack is one of the most critical malicious cyberattacks in the power system. This class of attack was first suggested by Liu et al.\cite{liu2011false}, in which the power system SE outputs are compromised by deliberately orchestrated injection of bad data into metre measurements. The theoretical frameworks for false data attacks are discussed in this section.  
\subsection{Stealthy FDI Attack} \label{stealthyy}
In a typical control center after SE is conducted, BDD techniques are employed to identify any injected bad data by computing residual vectors in terms of $\ell_2$-norm\footnote{$\ell_2$-norm of \textbf{r} is defined as \begin{math}
 ||\hat{\textbf{r}}||_2^2 = \sqrt{\sum {\hat {\textbf{r}}}^2}
\end{math}} between the original measurements $\mathbf{y}$ and the estimated measurements $\mathbf{\hat{y}=H\hat{x}}$, given by $\mathbf{||r||_2^2=||y-H\hat{x}||_2^2}$. However, research \cite{liu2011false} proved that BDDs are vulnerable to FDI anomalies. The outstanding feature of false data attacks is the residual vectors of the SE drop below the BDD's threshold despite the presence of maliciously corrupted measurements. Consequently, such strategically constructed false data attack vectors can bypass (i.e. remain stealthy in) the traditional BDD algorithms.
\subsubsection{FDI Attack Construction and Proof of Stealthiness}
In the presence of FDI attack, the adversary's goal is to introduce an attack vector \textbf{a} into the measurements without being noticed by the operator. Adversaries approach with different FDI attack strategies whereby the final effect of the malicious data results in compromising state variables across the power system domain. Generally, there are two main FDI anomaly construction strategies, one that requires knowledge of power system topology, and the other is based on a data-driven approach also known as the blind FDI attack strategy. Here, we use the former approach to demonstrate the stealthiness of the FDI attack. Let $\textbf{a}=[a_1, a_2,...,a_m]^T$ denotes the FDI attack, then measurements that contain this malicious data are represented by  
\begin{math}
 \textbf{y}_{false} = \textbf{y}+\textbf{a},  
\end{math} 
and 
\begin{math}
 \hat{\textbf{x}}_{false} = \hat{\textbf{x}}+\textbf{b}  
\end{math} refers to the estimated state vector after the FDI attack, where \begin{math}
  \textbf{b} = [b_1, b_2, ..., b_n]^T
\end{math} is the estimated error vector injected by the adversary. It is usually assumed \cite{liu2011false} that the attack vector \textbf{a} can be formulated as a linear combination of \textbf{H} given by \textbf{a} = \textbf{H}\textbf{b}. 

It has been proven \cite{liu2011false} that
if \begin{math}
 ||{\textbf{r}}||_2^2 < \tau
\end{math} it also holds true that \begin{math}
 \mathbf{||r_{\text{false}}||_2^2 } < \tau 
\end{math} for some detection threshold $\tau$. Hence, under \textbf{a} = \textbf{H}\textbf{b} the malicious measurement vector can pass the traditional BDD algorithms. 
\subsubsection{Sparsity of FDI Attack}
Usually \textbf{a} is assumed as a linear combination of the columns of \textbf{H} \cite{liu2011false}. However, the adversary's control can be limited to only over a few measurement devices. It could be because either the system has secure measurement devices which the attacker cannot access, or the attacker has limited physical access to the devices. This results in a sparse FDI attack \cite{liu2011false} \cite{ANWAR201758PCA4} \cite{ozay2013sparse}. FDI attack designed with only few non-zero components is called sparse attack and only small number of devices (let us say $k$) are required to launch the attack. Let the attack $\mathbf{\mathcal{A} = (a, \text{k})}$ contains the attack vector \textbf{a} and $k$ sets of compromised meters. Then, the sparse attack \cite{liu2011false} \cite{ANWAR201758PCA4} with $||\mathbf{a}||_0\leq k$ can be defined as $\ell_0$-norm minimization problem \cite{ozay2013sparse} and can be given as \begin{math}
\label{attk1Eg2}
\textbf{a}=
\begin{cases}
    \mathbf{Hb^\text{i}} , & \text{for i $\in$ \{1,...,k\}} \\
    \mathbf{0}, & \text{for i $\notin$ \{1,...,k\}}.\\
  \end{cases}
\end{math}
where the injected vector $\mathbf{b^\text{i}}$ is given by
\begin{math}
\label{bisV}
 \mathbf{b^\text{i}} \overset{\Delta}{=} [0,..,0,\underbrace{b_0}_\text{i}, 0,..,0]^T   
\end{math}
\subsection{Requirements for Stealthy FDI attacks} \label{reqFDI}
The requirements of FDI attacks are different from one application domain to the other. For instance, in wireless sensor networks (WSNs), the inherent wireless communication and broadcast channels among the nodes render more vulnerability to adversaries that may eavesdrop on all traffic, inject bad data reports containing erroneous sensor readings, or can even deplete the already limited energy capacity of sensor nodes \cite{1301328WSN}. In contrast, in the power system, it is difficult for an intruder to access the network parameters, hence, require more intelligent approach to launch a successful attack. In general, the FDI attacks create strong requirements from both the perspectives of the attackers and the system operators. The following are some of the main requirements for the FDI attacks under the cyber-physical Smart Grid environment.
\subsubsection{Rendering power system unobservability \cite{liu2011false}} Through the injection of false data an adversary can hijack and compromise the power system measurements which further results in the system unobservability. Typically, the attacker can remain undetectable at the control center while resulting in incorrect decisions of the state estimator. Even if the cyberattack can be detected by the SE, part of the power network may become unobservable where the SE cannot determine the system states.
\subsubsection{Partial-Parameter-Information} Earlier studies on the FDI attack models are based on the premise that the adversaries are capable of getting complete information of the power system topology. Authors in \cite{6503599DCincompleteInfo1} presented that it is also possible to construct stealthy attacks based on partial network information. However, attacks based on partial information require to satisfy the observability criteria. Another research direction ensures that the stealthiness (i.e. undetectability) of FDI attacks can also be modeled through data-driven or other partial-parameter-information approaches.
\subsubsection{Minimal Attack Vectors} For many reasons, the adversary's control can be limited to only over a few measurement devices. For this reason, stealthy FDI attacks should be designed with a very small attack magnitude and with only few non-zero components (i.e. attack sparsity) \cite{liu2011false} \cite{ANWAR201758PCA4}. Consequently, the attacker is required to compromise just smallest set of devices to cause network unobervability. 
\subsubsection{Attack Specificity} 
Whatever the motives of the cyber criminal are, the strategy behind the attack may be either indiscriminate  or targeted. The scope and impact of these two adversarial approaches are different. The indiscriminate attack may not require specific knowledge of the cyber-physical devices but launched arbitrarily against random Smart Grid elements. On the other hand, the targeted one can require a sophisticated approach which can be launched against targeted nodes. 
\subsubsection{Requirement on The Influence of The Attack} 
Attackers can approach in various ways to launch a successful attack and to cause a security risk on the Smart Grid. Some attackers want to exploit the data collected from sensors and networked devices across the power system. They may intend to exploit the weaknesses of sensors and communication protocols and launch the attack vector. Some typical examples of attack scenarios can be attack against sensor measurements (tampering power system parameter values in remote terminal units (RTUs) and PMUs). Another example can be by leveraging the communication protocols, where remote tripping injection can be performed by adversaries. In addition, attackers can infiltrate AMI-based communications networks in order to tamper with the contents of customer data that can result in disorder of the SE and other EMS functionalities. Others may intend to directly falsify the outcome of the state estimators \cite{liu2011false}.
\subsubsection{Requirement Based on Security Violations}
Some FDI-based malicious attackers try to infringe data availability, some violate data integrity, and others go against data confidentiality.  
\begin{enumerate} [label=(\alph*)]
\item Loss of data integrity:  For example, by injecting a systematically generated false data, a cyber intruder may compromise the integrity of the SE by hijacking a subset of metres and returning a modified data. The modification may involve deletion of data from the original meter readings, addition of bad data to sensor readings, or alteration of values in the hijacked measurements. The majority of FDI attacks, including, but not limited to, \cite{liu2011false} \cite{ANWAR201758PCA4} \cite{GIANI2014155} \cite{7885045} are based on this type of security violation.    

\item Loss of data availability: Furthermore, FDI attack can compromise the availability of critical information that is either intended to disrupt the power system or to stop its availability by shutting down network and communication devices  \cite{10.2015.066756}  \cite{7579185}
\item Attack on confidentiality: Although the effect of FDI on data confidentiality ranks among the least of all security objectives, the injection of false data could also violate the privacy of customers, especially in AMIs of the Smart Grid. This has become so common these days as illustrated in \cite{8918446}  \cite{chen2020blockchain}.
\end{enumerate}
\subsubsection{Requirements based on Attack Impact on The Power System} 
Threat actors can exploit Smart Grid security vulnerabilities that may lead to malfunctions in energy systems, operational failures in communications equipment as well as physical devices, and may even trigger a cascading failure. According to a report by NIST \cite{1213}, three potential impact levels, namely low, moderate, and high have been assessed for each of the Smart Grid security objectives following the degree of adversarial effect and associated risk level. 

Finally, the ultimate aim of FDI adversarial strategies is to pose significant consequences for the Smart Grid, such as causing sequential transmission line outages, maximizing operation cost of the system by injecting falsified vectors into subset of targeted meters, culminating in large-scale failure of the power system operation, and regional/national catastrophic impacts.
\section{Classification of FDI Attack Defence Strategies} \label{countermeasure}
The success of cyber-physical attacks in general and the FDI attack in particular depends on both the perspective of the adversary and the operator. In other words, it is highly likely that adversaries are subject to a trade-off between maximizing the probability of impact on various cyber-physical system components and minimizing the probability of detection of the launched attack. This section provides extensive review of existing state-of-the-art researches on the defence against the incumbent cyberattacks and mainly deals from the point of view of the power grid operator. The taxonomy is presented in section \ref{countermeasure}.  

There have been substantial research works on mitigation strategies against the FDI attack. We believe that taxonomy of the different countermeasures will help other researchers in the cyberattack defence arena to see correlations, differences, and to foresee future perspectives of these concepts. Here, we broadly classify the countermeasures into five categories with taxonomy depicted in Fig. \ref{taxCountermeasure} and details of each class are presented below.
\begin{figure*}[!ht]
\centerline{\includegraphics[width=140mm,scale=1.2]{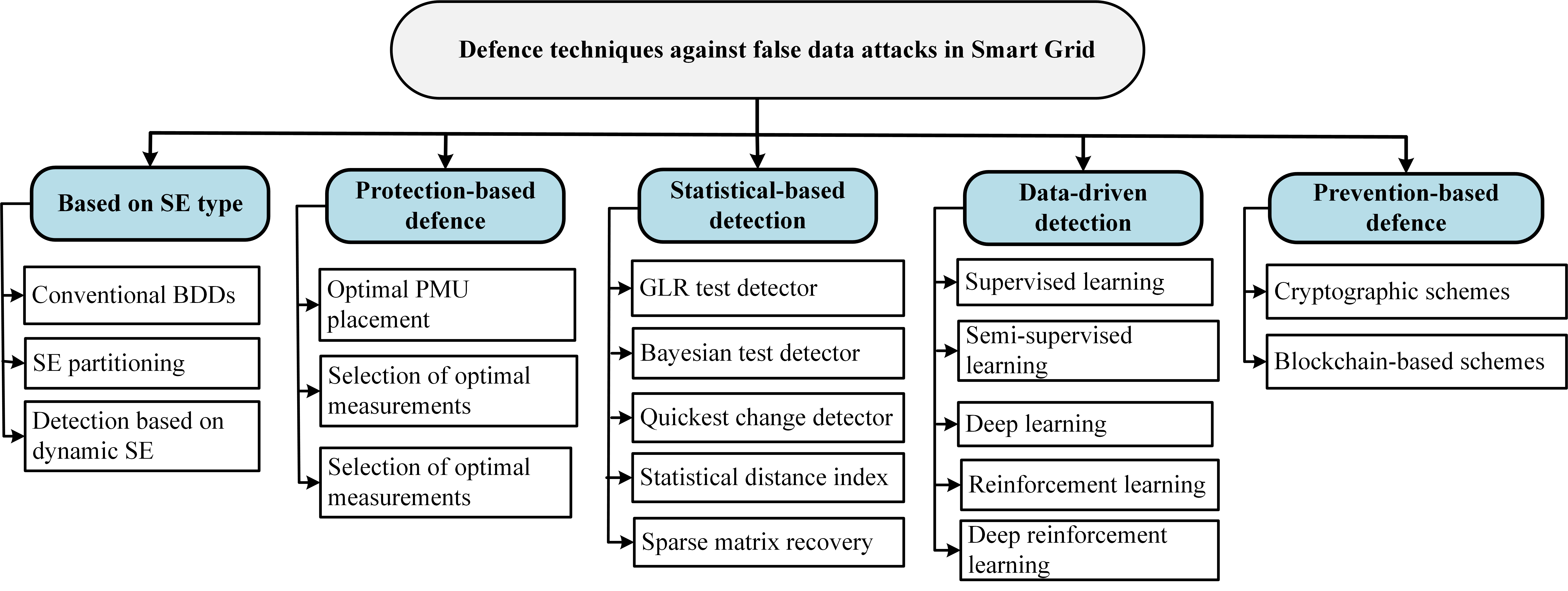}}
\caption{Taxonomy of the defence strategies}
\label{taxCountermeasure}
\end{figure*}
\subsection{Countermeasures Based on SE Type}
\subsubsection{Conventional Bad Data Detectors} \label{bdds}
BDD \cite{Abur} has been an
integral part of the power system state estimators that is used to detect and remove faulty measurements caused by error due to device malfunctions, communication channel problem, or cyberattacks, and is still widely used in various commercial EMS software. $\chi^2$ distribution test (or for short the $\chi^2-$detector) \cite{Abur} is the most widely used BDD in the power system SE. This technique employs hypothesis testing based on the WLS estimation to determine the cyber anomaly or bad data. The detail implementation of $\chi^2$ distribution test can be found in \cite{6655273}. 

Largest normalised residue (LNR) \cite{Abur} test is another metric for the identification of bad data. Given the residue between the observed measurement and the estimated measurement as $\mathbf{r=y-H\hat x}$, the LNR test is based on the largest or maximum value of the normalized residue $\mathbf{\hat {r}_\text{i}}$ for each measurement index $i$. 

Typically BDDs are used in the centralised state estimators. Few researches \cite{6655273} \cite{7868276} \cite{7867789} extended the use of BDDs in distributed based state estimators. In \cite{6655273}, they suggest to divide a power system into many non-overlapping subsystems according to the physical topology, apply SE and a $\chi^2$ distribution test for the detection of bad data in each subsystem. The findings reveal that the local degree of freedom is less than that obtained from the centralised SE, which results in a better identification of bad data. Similarly, in \cite{7868276} under distributed SE, a BDD is examined by taking into account the weight of local measurement residual of sub-areas of the power system and the overall change of measurement residual. A substation level BDD is also proposed in \cite{7867789}. However, the above BDD-based detection approaches didn't address FDI attacks (i.e. stealthiness and attack sparsity as discussed in Section \ref{stealthyy}), and thus are vulnerable to the FDI attack.
\subsubsection{Detection Based on SE Partitioning}
By decomposing the power grid (either physical partitioning as it is used in distributed SE or software partitioning) into many subsystems, measurement redundancy can be relatively minimised, and the threshold of false data in each subsystem can be smaller than the original system. As a consequence, the sensitivity of $\chi^2-$detector in each subsystem would increase, thus improving the chance of attack detection. For example, the authors of \cite{6567175} suggest adaptive graph partitioning for SE and applied $\chi^2-$detector.     
\subsubsection{Detection Based on Dynamic SE}
The absence of real-time information in the power system operation can indeed be attributed to its use of steady state estimators that produce input data for many EMS modules. Dynamic SE methods, on the other hand, model the time varying behaviour of the process, making it possible to predict the state variables ahead of time. In this case, the SE proves to be a great advantage for the system operator to conduct security analysis as well as other EMS functions. Kalman filter (KF) has been extensively utilised in dynamic SE. There are different extensions of KFs available, including extended KF, unscented KF (UKF), ensemble KF, particle filter which are designed for non-linear systems \cite{8624411}. The discussion of these techniques is beyond the scope of this paper and details of each dynamic SE is found in \cite{8624411}.

With the emergence of dynamic state estimators more efficient countermeasure strategies than the BDDs against the FDI attacks have been developed. In \cite{7084114}, a combined $\chi^2-$detector and cosine similarity matching techniques are employed for the detection of FDI attacks in Smart Grid where KF estimation have been used to measure any deviation from actual measurements. In the $\chi^2-$detector, the variation in the KF-estimated and measurements is used to identify the malicious attacks, and in the cosine similarity metric, the cosine of angle between the received measurements and the KF-estimated is computed to detect the attacks. While the $\chi^2-$detector has been confirmed as vulnerable to the FDI attack, the cosine similarity is found to have a better detection probability against the FDI attacks. Yet, the cosine similarity criterion is not efficient for sparse FDI attacks where the cosine angle between the received measurements and the injected data becomes almost unity, which bypasses the underlying detector. 

Similarly, in \cite{6897944}, the authors developed a detection of FDI attacks in Smart Grid, which is based on a KF-based state estimator. Euclidean distance detector was employed to detect the discrepancies between KF-estimated data and the received measurements. The Euclidean distance detector is used to quantify the difference between the estimated and observed states, where amplitude of the voltage signal is considered. If the difference is greater than a pre-determined threshold, the detector triggers a decision on whether attack exists or not. Although the proposed approach achieves better detection accuracy than the conventional BDDs, there are two drawbacks for this approach. First, it considers only time-invariant states where dynamic nature of the state variables is ignored. Moreover, the proposed detector cannot distinguish between FDI attack and a failure due to physical faults. Different from the above, reference \cite{vzivkovic2018detection} suggested the use of a combined UKF state prediction and WLS-based SE algorithm to detect inconsistencies between state vector estimates and, as a result, to detect false data attacks for non-linear measurement models. Normalised residual based on WLS and UKF estimates is computed, and compared to a predefined threshold. Although the combined WLS and UKF estimates has a better detection of the FDI than BDD and KF techniques, it has drawbacks. First, UKF state predictions are highly influenced by the non-linear transition matrix and system noise, which can potentially make it difficult to distinguish between attack-free and compromised states. Second, the accuracy of the detection relies on the UKF predicted outcome, whose uncertainties can result in high false positives. Third, a generalised FDI attack is considered, rather than a more stealthy and sparse FDI attack, which may pass the proposed detector. Therefore, data of various load forecasts, proper threshold selection, and threat model are among the critical points to consider for the robustness of the proposed methodology.  

Other FDI detection approaches based on the dynamic SE include: a spatio-temporal correlations \cite{7084114} among states of the power system,   short-term state forecasting-based approach for analysis of nodal state temporal correlations \cite{7313024}, and a graph signal processing-based \cite{8784391} scheme to determine the graph Fourier transformation of the estimated states and to filter the high-frequency components of the graph.
\subsection{Protection-Based Defence}
In Smart Grid cybersecurity, protection-based defence aims to deter the attacks by identifying a set of measurement devices and making them immune to the incumbent cyberattacks (e.g. using physical and efficient cryptographic methods) for ensuring observability of the states. The objective of introducing protections to components of the Smart Grid is that the attacker could not get enough measurements to start the FDI attacks, which otherwise will make the power system unobservable. The idea behind this defending technique is that, for a given grid topology, certain sensor readings affect more state vectors than others and should thus have a better cost-benefit ratio when secured through protection. Likewise, certain state vectors are reliant on more sensor data than others, and thus separately checking their estimation can restrict the ability of the hackers to exploit the sensor data without being noticed. Three main research approaches have been investigated: by deploying minimal number of PMUs, by selecting optimal set of measurements for protecting estimated state vectors, and by perturbing grid parameters, which are discussed as follows.  
\subsubsection{Optimal PMU Placement}
It has been found \cite{1717562} that the cyberattack protection capability of a power grid can be significantly enhanced with the integration of a few secure PMUs in the grid. This is because PMUs measure voltage and current phasors using a standard time source based on a global positioning system and therefore have the potential to provide precise time-stamped measurements for geographically distributed nodes. As a result, they have secured measurements, and are usually resilient against bad data injection attacks. For the same reason, in \cite{6666868}, linear programming based PMU placement algorithm is used to determine the number of PMU placements across a grid with $b$ number of branches and $n $ number of buses. For PMUs with $c$ current phasor measurements, they calculated $k=\binom{b}{c}$ possible combinations to assign those $c$ PMU configurations, and thus the number of possible PMU configurations for all buses is calculated as $N=\sum_{i=1}^{n} k_i$. Semidefinite programming approach \cite{7303980} has a better solution than \cite{6666868} for the problem of optimal placement of PMUs for protecting measurements against the malicious attacks. Similarly, mixed integer programming method \cite{GIANI2014155} determines the minimum number of PMUs needed to protect against unobservable data integrity attacks. 

Compared to a polynomial time-complexity of the linear programming and semidefinite programming, and an exponential time-complexity of mixed integer programming, greedy heuristics \cite{5751206Centralised3} can provide more optimal placement of secure PMUs to defend against the bad data injection attacks. Most of the strategies mentioned focus on evaluating the optimal placement of PMUs to enhance power system observability, cost, and protection, and improvement of SE. However, considering the adversary-operator dynamics, the adversary might have partial knowledge about the operator's corresponding defense measures, where they could optimize their attack strategy. For instance, the PMUs and the power system can be compromised by the adversary during device configuration process. Consequently, the aforementioned approaches are insufficient. As a solution to the drawback, \cite{9031416} proposes a predeployment PMU greedy algorithm against the attack where the most vulnerable buses are first secured and, then, a greedy-based algorithm is used to deploy other PMUs until the entire power system is observable. The defence space against the FDI attack can also be strengthened using a hybrid protection-based and detection-based scheme as suggested in \cite{wang2019two}, where the former is utilised to protect essential measurements from the intruder by means of physical defences, and the latter is used to identify modified data. They proposed a zero-sum static game-theoretic approach for the optimal deployment of the PMUs (for the PMU placement), and a false data identification and prediction based on historical patterns (for the detection). 

However, PMUs are very costly, and  it is not practical to install enough PMUs to secure sensor readings. It is definitely much more expensive especially with the emerging ubiquitous sensing infrastructure in to the large-scale Smart Grid. In addition, research has shown that PMUs are vulnerable to FDI attacks via GPS spoofing \cite{shepard2012evaluation}. Therefore, a more appealing security scheme is required to protect the power system against the FDI.
\begin{table*}[!ht]
\caption{Defence methods against FDI attacks in Smart Grid}
 \begin{adjustbox}
{totalheight={18cm},keepaspectratio,gstore totalheight=\mylength}
\begin{tabular}{|m{0.2cm}|m{4.5cm}|m{5cm}|m{5cm}|}
  \hline {\rotatebox{90}{\textbf{Category}}}
 & \textbf{Subcategory} & \textbf{Approaches/Algorithms} & \textbf{References} 
 \\ \hline
{\multirow{3}{*} {\rotatebox{90}{\textbf{Based on SE type}}}} & 
Conventional BDD & $\chi^2-$detector &  \cite{Abur}  \cite{7232283}  \\ \cline{3-4}
& & LNR detector & \cite{8231940} \cite{korres2011state} \cite{8352728} \\
  \cline{3-4}
& & Detection based on SE partitioning & \cite{6567175}  \\
  \cline{2-4}
  &  Detection based on dynamic SE & KF and extensions & \cite{6897944} \cite{vzivkovic2018detection} \cite{karimipour2017robust} \cite{chen2019novel}  \\   \cline{3-4}
& & Spatio-temporal correlations & \cite{7084114}  \\ \cline{3-4}
 & & State forecasting & \cite{7313024} \cite{7740014} \cite{8587473}  \\ \cline{3-4}
 & & Graph signal processing & \cite{8646454} \cite{hasnat2020detection} \cite{9163337} \\
 \hline {\multirow{8}{*} {\rotatebox{90}{\textbf{Protection-based defence}}}} 
  &  Optimal PMU placement & Integer linear programming & \cite{1717562} \cite{6666868} \\   \cline{3-4}
 & & Mixed integer semidefinite programming  & \cite{GIANI2014155} \cite{7303980} \\ \cline{3-4}  
 & & Greedy algorithm  & \cite{5751206Centralised3} \cite{7885045} \cite{8094240} \\ \cline{3-4}  
& & Predeployment PMU greedy  & \cite{9031416} \\ \cline{3-4} 
& & hybrid protection-detection  & \cite{wang2019two} \\ \cline{2-4}  
& Optimal measurement selection & Heuristic search (greedy algorithm and others) & \cite{7234893} \cite{bobba2010Protection1} \cite{7370787} \cite{5622046} \cite{6162362Protection2} \\ \cline{3-4}
& & Graph-theoretic &  \cite{6787030} 
\cite{9026334} \\  \cline{3-4}
  & & Game-theoretic & \cite{8219710} \cite{6378414} \cite{7570183} \cite{esmalifalak2013bad}\\ \cline{2-4}
 & Grid topology perturbation & MTD & \cite{6149267} \cite{6687991} \cite{9144465} \cite{rahman2014moving} \cite{8385215} \cite{8820088} \cite{8732433} \\  \cline{3-4} 
& & Hidden MTD & \cite{8252724} \cite{8762178} \cite{zhang2020hiddenness} \cite{8586470}\\ \hline 
{\multirow{10}{*} {\rotatebox{90}{\textbf{Statistical model}}}} &  GLR test detector & $\ell_1$-norm minimization &  \cite{6032057Centralised2} \cite{kosut2010malicious} \\  \cline{3-4}
& & Auto-regressive &  \cite{tang2016detection}  \\  \cline{2-4}
& Bayesian test detector & Game-theoretic &  \cite{mangalwedekar2017bayesian}  \\   \cline{3-4}
& & Joint estimation-detection & \cite{8101499} \cite{7127816} \cite{7086893} \\ \cline{2-4}

& Quickest change detection & CUSUM and adaptive CUSUM &  \cite{8278264} \cite{7587875} \cite{5766111} \cite{6949126} \\   \cline{3-4}
& & Sequential change detector & \cite{8808884} \cite{6982207} \cite{7896616} \cite{8485421} \\ \cline{2-4}
& Statistical distance & KL distance &  \cite{7035067} \cite{7961272} \\   \cline{3-4}
& & JS distance  &  \cite{8688077} \cite{8600016} \\ \cline{2-4}
& Low-rank and sparse matrix recovery & Sparse matrix optimization &  \cite{liu2013detection} \cite{6740901SparseOpt1} \\ \cline{3-4}
 & & Fast Go Decomposition & \cite{8489956} \\ \hline
{\multirow{14}{*} {\rotatebox{90}{\textbf{Data-driven}}}} 
&  Supervised ML & SVM & \cite{7063894} \cite{esmalifalak2014detecting} \cite{8555556} \\ \cline{3-4}
& & ANN & \cite{8221908} \cite{ganjkhani2019novel} \cite{8658084} \\ \cline{3-4}
& & KNN & \cite{7727361} \\ \cline{2-4}
&  Semi-supervised ML & semi-supervised ANN &  \cite{7063894} \cite{esmalifalak2014detecting}  \\ \cline{3-4}
& & Semi-supervised GMM & \cite{8221908} \\ \cline{2-4} & Deep learning & DFFNN &  \cite{8450487} \\   \cline{3-4}
& & CDBN &  \cite{7926429} \\ \cline{3-4}
& & DRNN & \cite{8334607} \cite{dehghani2020deep} \\ \cline{3-4}
& & CNN & \cite{8791598} \cite{9049087} \\ \cline{3-4}
& & GAN & \cite{9144530} \\ \cline{2-4}
& Reinforcement learning & Q-learning &  \cite{8248780} \\   \cline{3-4}
& & SARSA & \cite{8514804} \\ \cline{3-4}
& & Bayesian Bandit & \cite{8735819} \\ \cline{2-4}
&  Deep reinforcement learning & deep-Q-network &  \cite{8786811} \\
 \hline  {\multirow{5}{*} {\rotatebox{90}{\textbf{Prevention}}}}
& Cryptographic schemes & Encryption and dynamic Key management &  \cite{5622046} \cite{6303199} \cite{7571177}\\  \cline{3-4}
& & Authentication & \cite{hittini2020fdipp} \cite{mahmood2016lightweight} \cite{8854144}  \\ \cline{3-4}
& & end-to-end signature &  \cite{7822933} \\  \cline{2-4}
& Blockchain-based defence & Data protection &  \cite{8326530} \cite{mbarek2020enhanced} \\   \cline{3-4}
 & & Privacy preservation & \cite{8918446} \cite{chen2020blockchain}\\  \hline 
\end{tabular} \end{adjustbox}
\label{defTable}
 \end{table*}
\subsubsection{Protection Via Selection of Optimal Measurements} This is a security technique developed to defend SE against the injection of bad data through a carefully selected subset of measurements. For instance, reference \cite{bobba2010Protection1} employed a brute-force search for identifying optimal set of measurements and state vectors to ensure that stealthy data injection attacks are detected by the grid operator. The method enables the grid operator to choose a random number $q$ out of $n $ state variables, and to pick a random number $p$ out of $m$ sensors and should fulfill $\binom{m}{p}$*$\binom{n}{q}$ combinations for a given choice of $q$ and $p$, where $0 \leq q \leq n$ and $0 \leq p \leq m$. Similar to the brute-force method, fast greedy search algorithm \cite{7234893} can find optimal subset of measurements for protecting against the stealth FDI attacks. Further, by decomposing the connected elements of the power grid into many subnetworks, approximate solutions for the minimal number of measurements can be achieved, for example, using mixed integer linear programming \cite{7370787} model. 

In these three approaches, the system operator has to randomly select the number of measurements to be protected. Therefore, although the proposed method can be feasible for a small number of power systems, it is costly for a large-scale power grid. In contrast to \cite{bobba2010Protection1}, in \cite{5622046} protection measures are introduced, taking into account perfectly protected measurements (an ideal assumption that no stealth data injection attacks are possible) and non-perfectly protected measurements (where the operator seeks to maximize its protection level through some metric) considering the operator's budget as a constraint. In support of \cite{5622046}, \cite{6162362Protection2} derived exact and approximate solutions satisfying a protection criterion with a minimum number of measurement data points. 

However, determining such subset of measurements is a large-complexity problem. To alleviate these complexities, other approaches in this research direction include graph-theoretical and game-theoretical, both discussed below. 
\begin{enumerate}[label=(\alph*)]
 \item Protection Based on Graph-Theoretic: Graph-theoretic approaches are widely used for the power system observability analysis \cite{1216144}. They have also been used to define optimal protection problem to safeguard state variables with a minimal set of measurements. Some of the methods considered include the following:
\begin{itemize}
 \item Steiner tree-based graph theory \cite{6787030} (defending a set of priority-based critical state vectors).  
  \item Optimal and suboptimal solutions for state protections by modelling the Smart Grid as a minimum Steiner tree measurement problem \cite{9026334}.
\end{itemize}
\item Protection Based on Game-Theoretic: Game theories are important theoretical frameworks for the development of optimal decision-making of competing players, such as the adversary and the operator in the defending space of the Smart Grid.
\begin{itemize}
 \item The optimal set of protection can be formulated as a a three-level of defender-attacker-operator problem \cite{8219710} to deter the success of the coordinated attacks.   
 \item A  zero-sum Markov game-theoretical \cite{6378414} to model the defender-attacker relationships, where the defender can maximise their benefit by misleading the adversary to use incorrect cost functions of the grid.
\item Adaptive Markov \cite{7570183} technique to dynamically compute an optimal defense scheme against malicious attackers with  dynamic and unpredictable behavior.
\end{itemize}
\end{enumerate}
\subsubsection{Grid Topology Perturbation}
Most of legacy IT systems are static, adopted for simplicity over time. However, in a static system, hackers can have enough time for reconnaissance against the system, enough opportunity to learn the flaws and related attack vectors, and ultimately, to initiate attacks against the system. Recently, moving target defense (MTD) \cite{jajodia2011moving} has emerged as a proactive defence strategy that has been studied in various areas of cybersecurity. MTD is helpful to maximise the complexity against adversaries by implementing uncertainty, or to increase the cost attack. 

Similarly, MTD has become popular among grid operators for deceiving adversaries. Grid operators can proactively protect the measurements against the malicious attackers by introducing perturbations to network data or topology. The key purpose of this strategy is to defeat the malicious user who presumably knows network data or topology configurations. The perturbation can be done by systematically changing system settings that
adversaries might need to aim for launching their attacks, in order to nullify their prior information of the system and making it impossible for the adversaries to adapt their attack space. In this regard, as the topology perturbation patterns are hidden from the hackers, they cannot compute and generate the proper response for the measurements or topology under their control that makes the FDI attack unable to correct to remain undetectable. 

There are different kinds of perturbations for protecting key grid elements against the FDI attacks. In \cite{6149267}, for example, the authors applied impedance changes through a key space approach to a number of selected transmission lines by leveraging D-FACTS\footnote{distributed flexible AC transmission system (D-FACTS) are devices installed on power line to change the power flow by altering impedance of the line} devices in order to generate noticeable system changes that the adversary cannot foresee. The anticipated system response is predicted and compared to the observed measurements. Nonetheless, if a perturbation sequence has been implemented \cite{6149267} in such a way that the system is made to revert to a previously observed state, the difference between the anticipated result of the perturbation and the actual result of the probe would reveal the presence of false data. MTD can also utilise both a randomized set of measurements considered in SE and the topology of transmission line \cite{rahman2014moving}. Similar line of researches include \cite{6687991} \cite{9144465}. 
 
However, the above-mentioned MTD strategies have been implemented under a weak adversarial environment in which they overlook the likelihood that sophisticated FDI attackers may also attempt to identify MTD changes before they execute the attack. As a remedy for this limitation, the authors of \cite{8252724} introduce a hidden MTD, an approach that hardens the stealthiness of the MTD. Similar researches have been conducted in this category including \cite{8762178}.  
\subsection{Detection Based on Statistical Modelling}
Statistical models, most of which started to take hold about two centuries ago, are still widely used in a number of modern-day fields. Several research efforts of statistical-based detection frameworks against falsified injection of data have been addressed by the Smart Grid community. These approaches are summarised into Generalized likelihood ratio (GLR) test detector, Bayesian test framework, quickest change detection, statistical distance index, and sparse matrix recovery. 
\subsubsection{GLR Test Detector}
GLR test detector is one of such statistical models used for detecting cyberattacks in the power system by leveraging the likelihood ratio of statistical tests. While it is usually not feasible to use the GLR test detector to detect a large number of compromised samples, it can do well to detect weak FDI attacks \cite{6032057Centralised2}, where $\ell_1$-norm minimization is proposed to solve the detection problem. In particular, it has been noted in \cite{6032057Centralised2} that if multiple measurement samples are available under the same sparse FDI attack, the GLR test detector can be asymptotically optimal in the sense that gives a very low probability of miss detection. Although the FDI detector in \cite{6032057Centralised2} is valid under AWGN distribution, a study \cite{tang2016detection} has shown that it doesn't satisfy when the measurement are corrupted by non-Gaussian \cite{ANWAR201758PCA4} noise distributions. The authors of \cite{tang2016detection} used an independent component analysis along with the GLR test detector for an FDI attack on the basis that the power system measurements are subject to a colored Gaussian noise (modeled through auto-regressive process). \subsubsection{Bayesian Test Detector}
Bayesian-based statistical frameworks are essential for decision-making by leveraging prior knowledge and new evidence. For example, a strategic attacker-defender Bayesian game-theoretic detection technique \cite{mangalwedekar2017bayesian} against FDI may be established where the Bayesian game is played on each node in the event of an attack on that node and a critical set of measurements to be defended is obtained for the particular node. Further, in \cite{8101499}, a Bayesian-based detector has been proposed for each monitoring node using a distributed architecture in WAMS. Once the probability of FDI attack vectors is determined by Bayesian inference, then a recursive Bayesian-based prediction is derived for the attack detection using measurements obtained from real power transmission grid and simulated measurements. Other related works of Bayes approach for the detection of FDI attacks include \cite{7127816} \cite{7086893} \cite{5464816Centralised4}.
\subsubsection{Quickest Change Detection}
Quickest change detection (QCD) \cite{poor2008quickest} (which can be performed close to the real-time detection) is a mechanism to detect sudden changes as soon as possible on the basis of sequential or real-time observations in such a way that minimizes the lag between the moment a change appears and the time it is observed. When distributions of before and after change are explicitly defined, a variety of detection methods have been suggested under different conditions. Unlike the static BDD detection procedures, which are based on a single measurement at a time, the QCDs consider use of dynamic change detection procedures. Overall, the objective of this approach is to minimise the average detection time under certain detection accuracy limitations. QCD-based detection techniques \cite{poor2008quickest} \cite{5766111} can be used with Bayesian model, Non-Bayesian model (e.g. CUmulative SUM (CUSUM), adaptive CUSUM test), and statistical hypotheses test. 

The following describes the literature that utilise the QCD technique to detect FDI attacks in smart Grid:
\begin{itemize}
\item A Markov-chain-based QCD algorithm for dynamic SE is proposed to detect and remove FDI attacks \cite{8808884}, \item a joint dynamic CUSUM and static $\chi^2-$detector in which the former leverages historical states and the latter utilises a single measurement at a time \cite{7587875}, 
\item generalized CUSUM algorithm is suggested for quickest detection of FDI attacks for dynamic KF-based state estimator under centralized and distributed settings \cite{8278264}.
\item adaptive CUSUM methodology for quickest change detection using a linear unknown parameter solver \cite{5766111},
\item A Markov-chain-based adaptive CUSUM for a real-ti,e detection of FDI attacks \cite{6949126}, 
\item sequential detection of centralized and distributed FDI attacks based on the GLR test \cite{6982207},
\item generalized sequential likelihood ratio test for a decentralised system \cite{7896616}.
\end{itemize}
\subsubsection{Detection Based on Statistical Distance Index}
A statistical distance quantifies the consistency of two probability distributions through, for example, a variational distance between the distributions. Kullback–Leibler (KL) distance \cite{7035067} and Jensen-Shannon (JS) distance \cite{8688077} have recently been used for detecting malicious power system measurements by calculating the dissimilarity among  probability distributions obtained from measurement variations. KL distance metric has been suggested \cite{7035067} \cite{7961272} to track the measurement dynamics and to detect FDI attacks. When bad data is injected into the power systems, the variations in the probability distributions of the measurements deviate from historical data, leading to a greater distance of KL. Likewise, the JS distance-based detection framework \cite{8688077} monitors dynamics of probability distributions obtained from historical measurement variations and real-time measurement variations. Similarly, the JS distance based detection is proposed to detect FDI of electricity theft in AMI \cite{8600016}. Similar statistical distance-based approaches have also been used along with data-driven techniques (see Section \ref{dataDriven}).
\subsubsection{Detection As Low-Rank And Sparse Matrix Recovery}
Another interesting research to investigate is the detection of a low-sparse FDI attack. In addition, measurement matrix obtained at the control center has low dimensional structure due to the inherent temporal correlation of states of the power gird. Taking into account the low-rank structure of the measurements and the low-sparse of the false data attacks, low-rank and sparse matrix recovery, an approach which has found applications in various fields, is another alternative for the defence against the incumbent cyberattacks. 
 
The detection problem of measurements with FDI attack have been formulated as a low-rank and sparse matrix recovery \cite{liu2013detection} \cite{6740901SparseOpt1} \cite{8489956}. Liu et. al \cite{6740901SparseOpt1} formulate the problem of detecting FDI attacks as low-rank matrix recovery in the form of augmented nuclear norm\footnote{nuclear norm is a convex optimization problem that is used to search for low-rank matrices.} and $\ell_1$-norm minimization solved through. By considering the intrinsic low-dimensional structure of temporal attack-free measurements of power grid and sparse FDI malicious attacks, they extended their work in \cite{liu2013detection} to a problem of sparse matrix optimization in \cite{6740901SparseOpt1}, solved using low-rank matrix factorization. On the other hand, while the results of \cite{liu2013detection} and \cite{6740901SparseOpt1} has a good computational efficiency, they have quite low FDI detection accuracy. Therefore, in order to obtain a better balance between the detection accuracy and the computational performance, the authors of \cite{8489956} proposed a new approach known as 'Fast Go Decomposition' considering the low rank behaviour of the measurement data and the sparse FDI attack.
\subsection{Data-Driven Attack Detection} \label{dataDriven}
Various ML techniques have been employed for the detection of FDI attacks in smart power grid. Supervised learning classifiers \cite{7063894} \cite{esmalifalak2014detecting} \cite{8555556} \cite{8221908}
\cite{ganjkhani2019novel} \cite{8658084} \cite{7727361} are the most popular ML techniques for the detection of false data. These techniques can reflect the statistical characteristics of the power system using historical data and may allow the training model a better decision if redundant power system measurements are available. Historical training data can include class labels of normal versus tampered and using such training data a new observation is predicted as either false data injected or normal data. Ozay et al \cite{7063894} suggested supervised learning-based binary classifiers using statistical deviations between the FDI-corrupted and secured measurements. Similarly, in \cite{esmalifalak2014detecting} an FDI detection is used based on principal component analysis for reducing dimensionality of measurement data and supervised learning over labeled data for the classification. In the literature, various supervised ML algorithms are employed including support vector machine (SVM) (e.g. in \cite{7063894}, \cite{esmalifalak2014detecting}, \cite{8555556}), artificial neural network (ANN) (e.g. in \cite{ganjkhani2019novel}, \cite{8221908}, and  \cite{8658084}), k-nearest neighbor (KNN) (e.g. in \cite{7727361}). 

One of the disadvantages of supervised learning techniques, however, is that they require far more labelled data, which is often difficult to obtain. For this reason, semi-supervised learning techniques address the problem of supervised learning by using partially labelled samples. In other words, this approach seeks to label unlabeled data points using information gained from a limited number of labeled data points. References \cite{7063894} \cite{esmalifalak2014detecting} \cite{8221908} also employed semi-supervised algorithms. Attack strength and sparsity are the two main factors that should be considered in the detection frameworks. While most proposed supervised and semi-supervised approaches consider a relatively high magnitude of FDI attacks, their detection accuracy is low for a very small attack magnitudes. On the other hand, deep learning (DL) techniques can extract high-dimensional temporal features of the FDI attacks with historical measurement data and can use the known features to detect various magnitudes of FDI attacks in real-time. More specifically, the latest advance in graphics processing units (GPU) computation provides the basis for deep neural networks such as deep feedforward neural network (DFFNN) \cite{8450487}, deep belief network (DBN) as used in \cite{7926429}, deep recurrent neural network (DRNN) \cite{8334607} and \cite{8334607}, convolutional neural network (CNN) \cite{8791598}, and a semi-supervised deep learning approach using generative adversarial network (GAN) framework \cite{9144530}.   
\subsection{Prevention-Based Defence} \label{prevDef}
Intelligent and integrated cyber-physical resources intended to improve the stability and reliability of the Smart Grid could be used as weapons against the grid itself. Without proper cyberattack prevention schemes, the Smart Grid can be more vulnerable especially when it is connected to the Internet via less secure wireless communication systems such as ZigBee \cite{7997893} \cite{reda2018application} and Wi-Fi. Until malicious hackers successfully launch their attack vectors and inflict irreparable damage to the power grid, they typically proceed through comprehensive technical stages, such as the reconnaissance for investigating the technical flaws of the system. However, most of the defence countermeasures are based on identification of the false data attack normally after the threat compromised the data integrity at the control center, during the transmission, or at measurement devices. To this end, lack of adequate preventive security measures against coordinated false data attacks could be disastrous. Hence, as part of Smart Grid cybersecurity, preventive security measures are essential in the battle against attacks such as the FDI. By providing prevention schemes across key cyber-physical resources, we can deter the malicious users against  unauthorised access of EMS/DEM/MMS critical OT database systems, exploitation of the communication protocols (e.g IEC 61850), compromising user privacy or data integrity via smart meters, and tampering IEDs or interception of data transmission in WSN, IoT, and cognitive radio \cite{reda2019firefly}.

In order to respond effectively to threats, it is necessary to implement a wide variety of cyberattack prevention and security techniques across the Smart Grid. Further, effective cybersecurity can also be accomplished by combining both preventive and detective systems. In this survey paper, the most prominent prevention systems in Smart Grid, such as cryptographic schemes and privacy preservation using Blockchain, have been summarised, which are discussed below. 
\subsubsection{Cryptographic Schemes}
It is highly likely that adversaries can exploit communication channels when measurements are sent from sensors to control centers or when customer data is transmitted from smart metres to the control centres over unencrypted communication channels. For example, the unencrypted communication channel of plain text transmission over the SCADA network or the IEC 61850-SAS compliant communication protocol could be hacked by cyber-enabled malicious actors, which could further mislead the control centre and other consequential impacts. Cryptographic techniques are well matured and over the years they have been applied in various domains for preventing different cyberattacks. However, cryptographic protocols are difficult to implement in the Smart Grid subject to the limited computational capabilities and the deployment in hostile environments of the measurement sensors or related devices. Therefore, fast and efficient cryptographic operations are required for implementation in the Smart Grid to guarantee the accuracy and integrity of measurements against FDI attacks. 

In \cite{5622046}, it was suggested that encrypting sufficient number of IEDs could improve measurement protection against stealthy FDI attacks and could increase overall system security in the utility control centre. Dynamic key management-based cryptographic protocols can also deter cyberattacks against privacy in Smart Grid wireless communication networks \cite{6303199}. Similarly, a dynamic and periodic secret-key generation scheme over Smart Grid communication network against various cyberthreats including the FDI is proposed in \cite{7571177}, which enables a resilience so no adversary can exploit the network over a longer period of time even if they know a secret key. 

The Smart Grid infrastructure involves millions of electronic devices that link customers to different cyber-physical entities. This calls for the need of a strict authentication process, which is vital for the verification of the customers and the devices. For example, strict authentication schemes can be implemented in IEC 62351 EMS-compliant security standard. For Smart Grid distribution systems, an FDI prevention protocol is proposed in \cite{hittini2020fdipp} that focuses on data integrity by preventing packet injection, replication, modification, and access to rogue nodes for the IEC 61850-90-1 SAS communication security standard. In particular, three stages accompany the operation of their proposed protocol: node authentication (authentication techniques across the distribution network including routers, gateways, inter-substation devices), peer authentication (authentication of a routing protocol when using cloud platform for the distribution system), and data transmission. Lightweight hash-based message authentication protocols \cite{mahmood2016lightweight} \cite{hittini2020fdipp} are also critical for thwarting false data attacks in IP-based data transmission in the Smart Grid environment. Further, a lightweight authentication scheme \cite{8854144} with reduced energy, communication, and computational overheads can establish a secure communication between two communicating parties, such as smart meters and wireless base stations, and can provide energy efficiency in a resource-constrained environment 

Other prevention methods include end-to-end signature schemes, which can protect data during an end-to-end communication in Smart Grid. For example, these schemes can protect legitimate commands transmitted from the control center to IEDs against malicious commands sent by adversaries \cite{7822933}. 
\subsubsection{Blockchain-Based Defence}
Blockchain-based prevention schemes can strengthen the ability of the Smart Grid to protect itself from against the incumbent cyberattacks. For example, data protection capabilities of Smart Grids against FDI attacks can be harnessed by introducing a distributed Blockchain-based reconfigurable SCADA network features for geographically distributed sensors \cite{8326530}. Here, during transmission or reception, each information in the distributed Blockchain network is cryptographically connected block by block, and includes signatures for verification. Further, Blockchain can be used to preserve the privacy of user's energy data against coordinated data integrity attacks. A distributed blockchain network based data management on mobile nodes for the microgrids trading is proposed in \cite{mbarek2020enhanced} that aims the prevention against false data attacks.

Blockchain-based privacy preservation mechanisms are used to protect network nodes or data transactions in the form of a peer-to-peer crypto connectivity \cite{8326530}. In \cite{8918446}, a Blockchain-based bi-level privacy module and anomaly detecting module is designed to verify data integrity and mitigate attacks of false data. A variational autoencoder and anomaly detector is proposed, where the former is applied for transforming data into an encoded format for preventing the cyberattacks and the latter is used to detect any interference attack.
    \section{Literature Review Method}\label{revMethod}
The method of literature review represents the foundational first step that makes up the skeleton of the knowledge base and largely dictates its reconstruction in the successive analysis of the literature. Therefore, the process of a systematic search, selection, analysis, and critical evaluation of the literature is described in this section.
\begin{figure}[!ht]
	\centerline{\includegraphics[width=60mm, scale=0.5]{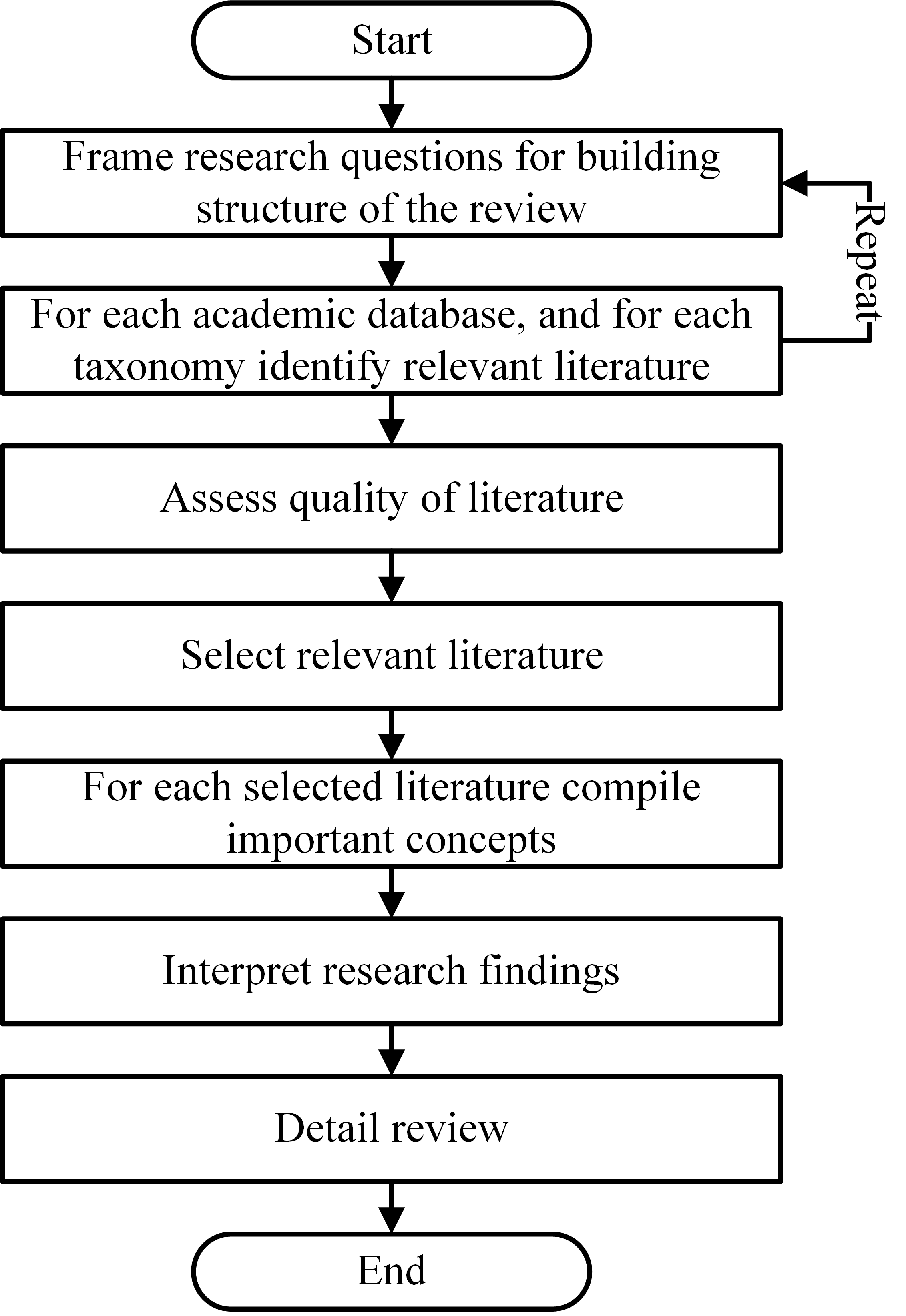}}
	\caption{Literature review methodology}
	\label{liteMethod}
\end{figure}
\subsection{Literature Search Methodology}
It seems that the literature search process plays an important role in crafting a comprehensive analysis of a topic. The literature survey of this paper is based on the search methodology adopted by Webster and Watson \cite{webster2002analyzing}. The systematic identification of high-quality publications (namely review articles, journals, conferences, and Books), technical reports, and dissertations are reflections of the correct selection of databases, keywords, the time covered, the papers considered in the literature search, and performing backward and forward searches \cite{alex213419}. 

Fig. \ref{liteMethod} is a description of the methodology used for literature search on this paper. The following academic research databases are considered: IEEE Xplore (IEEE/IET) digital library, Elsevier ScienceDirect, Association for Computing Machinery (ACM) digital library,  SpringerLink, and Others.
   
To find relevant papers, the flow chart of Fig. \ref{liteMethod} is applied for each of the above academic research databases. Using the first step, keywords using Google Scholar and Microsoft Academic were identified with respect to the defence countermeasures. "Smart Grid", "power system", "false data injection", and "cyber security" are common keywords used along with "detection", "defence", "mitigation", and "countermeasure". 
\subsection{Literature Selection and Analysis}
Primarily, we reflect entirely on the defence of FDI threats with respect to the Smart Grid cybersecurity, as there are also FDI articles related to other areas such as WSN, healthcare, software-defined networks, and so on. Another consideration is, while all the scholarly research sources considered are prestigious and are assumed to publish quality works, further evaluations were made using scientific journal ranking platforms to assess quality of the journals and the CORE\footnote{CORE: Computing Research and Education Association of Australasia (https://www.core.edu.au/)} was used for the conferences. Based on the search method as described above, a systematic literature selection and analysis are used which are described here. First, aggressive search was conducted using the above keywords and step 2 of Fig. \ref{liteMethod} that resulted in abundant number of papers. Then, after a systematic refinement across the subcategories of the taxonomy of the FDI attack mitigation techniques, relevant literature were selected. In addition to the keywords, titles and abstracts were considered for correctly categorising the selected papers. Next, important concepts were assembled for each of the chosen articles, accompanied by an overview of research results, and a thorough analysis. After an in-depth analysis of the literature, approximately 111 papers are found which, to varying degrees, dealt with the topic of the defence for FDI attack in Smart Grid cybersecurity. Note that the study of FDI attack in Smart Grid started in the late 2009. Therefore, the search for the most relevant literature of our survey starts from 2009 up to October 30, 2020 although related literature such as the BDD goes back in time before 2009. Table \ref{pubLit} is a summary of the number and source of the relevant publications considered in our survey paper. 
\begin{table}[ht]
 \caption{Summary of relevant publications}
 \centering
 \begin{tabular}{|m{2cm}|m{2.5cm}|m{1.5cm}|m{1.5cm}|m{1.8cm}|m{1.5cm}|}
   \hline 
\textbf{Database source} & \textbf{No. of relevant papers} & \textbf{Survey articles} & \textbf{Orig. res. articles} & \textbf{Conferences} & \textbf{Books/ Thesis} \\  \hline
IEEE Xplore & 91 & 4 & 59  & 28 & 0\\  \hline
Elsevier SD & 7 & 2  & 5  & 0 & 0\\ \hline
ACM & 3 & 1 & 1 & 1 & 0\\ \hline
Springer & 5 & 1  & 2  & 1 & 1 \\ \hline
Others & 5 & 0  & 3 & 2 & 0 \\ \hline
\textbf{Total} & \textbf{111} & \textbf{8}  & \textbf{70}  & \textbf{32} & \textbf{1} \\ \hline \end{tabular}
\label{pubLit}
\end{table}
\begin{table}[ht]
 \caption{Evaluation criteria for the defence strategies against FDI attacks in Smart Grid}
 \centering
 \begin{tabular}{|m{3cm}|m{10cm}|}
 \hline 
 \textbf{Criterion} & \textbf{Description} \\ 
 \hline 
Attack model &  Review the countermeasures from the point of considered attack models \\ \hline 
  Power flow model &  Adversaries use different approaches with different power flow models, so countermeasures are reviewed and compared accordingly  \\ \hline
  Defence algorithm & Review various defence techniques studied in the literature   \\ \hline
   Network architecture & Relevant articles are reviewed from  network-centric point of view  \\ \hline
   Attack target & Articles are compared on the basis of the attack target     \\ \hline
   Performance metric & Show the main claim of the research exemplifying the performance  \\ \hline
    Experimental platform &  Show the theoretical proofs or hardware testbeds utilized to justify the method \\  \hline 
    \end{tabular}
    \label{criteriaTable}
\end{table}
\begin{table*}
 \caption{Comparison of defence strategies against FDI attacks in Smart Grid cybersecurity}
 \label{cmpTable}
 \begin{adjustbox}
{totalheight={18.5cm},keepaspectratio,gstore totalheight=\mylength} 
\begin{tabular}{|m{0.15cm}|m{0.15cm}|m{2.5cm}|m{1.5cm}|m{0.10cm}m{0.10cm} m{0.10cm} m{0.10cm}m{0.10cm}|m{0.1cm}m{0.1cm}m{0.1cm}m{0.1cm} m{0.7cm}m{0.1cm}m{0.1cm}|m{2.5cm}|m{1cm}m{0.1cm}m{0.1cm}|}
 \hline
  & & & & \multicolumn{5}{c|}{\textbf{Attack model}}& \multicolumn{7}{c|}{\textbf{Attack target}} & & \multicolumn{3}{c|}{\textbf{Exp. platform}} \\      
 {\rotatebox{90}{\textbf{Category}}} & {\rotatebox{90}{\textbf{Subcategory}}} & \textbf{Algorithm} & \textbf{Reference} & {\rotatebox{90}{Complete information}}  & {\rotatebox{90}{Partial information}} & {\rotatebox{90}{LR attack}} & {\rotatebox{90}{GT attack}} & {\rotatebox{90}{Data-driven}} & {\rotatebox{90}{EMS}} & {\rotatebox{90}{AGC}} & {\rotatebox{90}{DEM}} & {\rotatebox{90}{MMS}} & {\rotatebox{90}{Network comm.}} & {\rotatebox{90}{Intelligent device}} & {\rotatebox{90}{Renewable DER}} & \textbf{Perf. metric} & {\rotatebox{90}{Bus system}}  &
   {\rotatebox{90}{Simulation}} & {\rotatebox{90}{Test bed}} \\   \hline
{\multirow{28}{*} {\rotatebox{90}{\textbf{Based on SE type}}}} & {\multirow{4}{*} {\rotatebox{90}{Conventional BDD}}} & $\chi^2$-detector & \cite{Abur}$^{A,c}$ & & & & & & $\checkmark$ &  &  & &  & & & DR vs $\tau$ & 3 & $\checkmark$ &    \\  \cline{4-20}
 & & & \cite{7232283}$^{D,c}$ & & & & & & $\checkmark$ &  &  & & SCADA & & & Estimated error vs $\tau$ & 265 & $\checkmark$ & \\\cline{3-20}
& & LNR detector & \cite{8231940}$^{A,c}$ & & & & & & $\checkmark$ &  &  & & PMU & & & Normalised residue & 30 & $\checkmark$ &  \\  \cline{4-20} 
& & & \cite{korres2011state}$^{A,c}$ & & & & & & $\checkmark$ &  &  & & PMU & & & Normalised residue vs $\tau$, PE & 14 & $\checkmark$ & \\  \cline{4-20}
& & & \cite{8352728}$^{A,c}$ & & & & & & $\checkmark$ &  &  & & SCADA & & & Normalised residue vs gross error & 30 & $\checkmark$ &  \\  \cline{3-20}
& & SE partitioning & \cite{6567175}$^{A,c}$ &$\checkmark$ & & & & & $\checkmark$ &  &  & & SCADA & & & DR vs $\tau$ & 39 & $\checkmark$ &  \\  \cline{2-20}
& {\multirow{2}{*} {\rotatebox{90}{Detection based on dynamic}}} & KF \& extensions & \cite{6897944}$^{A,c}$ &$\checkmark$ & & & & & $\checkmark$ &  &  & & AMI & & & DR, FAR & 9 & $\checkmark$ &    \\  \cline{4-20}
& & & \cite{vzivkovic2018detection}$^{A,c}$ &$\checkmark$ & & & & & $\checkmark$&  &  & & SCADA, PMU & $\checkmark$ & & MAPE, FDI & 14, 300 & $\checkmark$ &  \\ \cline{4-20}
& & & \cite{karimipour2017robust}$^{A,c}$ &$\checkmark$ & & & & & $\checkmark$&  &  & & SCADA & $\checkmark$ & & DR, FAR & 118 & $\checkmark$ &   \\  \cline{4-20}
& & & \cite{chen2019novel}$^{A,c}$ &$\checkmark$ & & & & & $\checkmark$&  &  & & SCADA & $\checkmark$ & & DR, $\tau$ & 14 & $\checkmark$ &   \\  \cline{3-20}
& & Spatio-temporal correlations & \cite{7084114}$^{D,cd,RL}$ & & &$\checkmark$ & & & $\checkmark$ &  & $\checkmark$ &$\checkmark$ & AMI & $\checkmark$&$\checkmark$ & FPDR, Outage rate, Energy cost, FDI & - & $\checkmark$ &  \\  \cline{3-20}
& & State forecasting & \cite{7313024}$^{D,c}$ & $\checkmark$& & & & & $\checkmark$ &  &  & & SCADA, PMU & & &DR, FAR, \# of PMUs & 14, 118 & $\checkmark$ & \\  \cline{4-20}
& & & \cite{7740014}$^{A,cd}$ & & & &$\checkmark$ & & $\checkmark$ &  &  & & SCADA & & & residual changes & 14 & $\checkmark$ &  \\  \cline{4-20} 
& & & \cite{8587473}$_{SR}^{D,c}$ & & $\checkmark$& & & & $\checkmark$ &  &  & & AMI & & &F1, SNR, \# of features & 14 & $\checkmark$ &  \\  \cline{3-20}
& & Signal processing & \cite{8646454}$^{D,c}$ & & &$\checkmark$ & & & $\checkmark$ &  &  & & PMU & & & DR vs phase/mag. deviation &14 & $\checkmark$ &   \\  \cline{4-20}
& & & \cite{hasnat2020detection}$^{A,c,RL}$ & & &$\checkmark$ & & & $\checkmark$ &  &  & & SCADA & & & DR, FAR &118 & $\checkmark$ &  \\  \cline{4-20} 
& & & \cite{9163337}$^{A,d}$ & & &$\checkmark$ & & & $\checkmark$ &  &  & & SCADA & & & MSE & 8 & $\checkmark$ &  \\ \hline {\multirow{22}{*} {\rotatebox{90}{\textbf{Protection-based defence}}}} &
{\multirow{8}{*} {\rotatebox{90}{Optimal PMU placement}}}  & Integer LP & \cite{1717562}$^{D,c}$ & && & & & $\checkmark$ &  &  & & SCADA & & & DR&57, 118 & $\checkmark$ &  \\  \cline{4-20}
  & & & \cite{6666868}$^{D,c}$ & && & & & $\checkmark$ &  &  & & SCADA & & & MSE &14 & $\checkmark$ &   \\ \cline{3-20}
& & MISDP & \cite{GIANI2014155}$^{A,c}$ & $\checkmark$&& & & & $\checkmark$ &  &  & & SCADA & & & Cost of undetected attack vs \# of PMUs & multiple & $\checkmark$ &  \\  \cline{3-20}
& & Greedy algorithm & \cite{5751206Centralised3}$_{SR}^{D,c}$ & &$\checkmark$& & & & $\checkmark$ & & & &SCADA & & & Subset of meters protection & multiple & $\checkmark$ &  \\  \cline{4-20}
&  & & \cite{7885045}$_{SR}^{D,c}$ & &$\checkmark$& & & & $\checkmark$ & & & & SCADA& & & Attack cost vs PMU placement &multiple & $\checkmark$&   \\  \cline{4-20}
&  & & \cite{8094240}$^{D,c}$ & &$\checkmark$& & & & $\checkmark$ &  &  & & SCADA & & & SE error deviation vs PMU placement & multiple & $\checkmark$ &  \\  \cline{4-20} 
& & & \cite{9031416}$^{A,c}$ & &$\checkmark$& & & & $\checkmark$ &  &  & & SCADA & & & PMU placement vs attack cost, time overhead & 9,14,30 & $\checkmark$ &    \\ \cline{3-20}
& & Hybrid & \cite{wang2019two}$^{A,c}$ & &$\checkmark$&$\checkmark$ & & & $\checkmark$ & $\checkmark$ &  & & SCADA & & & Defence probability vs nodes &14 & $\checkmark$ & \\  \cline{2-20}
& &Graph-theoretic & \cite{6787030}$^{D,c}$ & $\checkmark$&& & & & $\checkmark$ &  &  & & SCADA &$\checkmark$ & & Optimal meter protections &14, 57, 118 & $\checkmark$ &  \\  \cline{4-20}
& & &\cite{9026334}$^{D,c}$ & &$\checkmark$& & & & $\checkmark$ &  &  & & SCADA & & & Optimal meter protections &30, 57, 118 & $\checkmark$ &  \\ \hline
\end{tabular} \end{adjustbox}
{\fontsize{6.5}{9}\selectfont 
\begin{tablenotes}
\item $[Ref^{D/A}]$: DC/AC model, $[Ref^{c/d}]$: centralised/decentralised architecture, $[Ref^{cd}]$: centralised and decentralised architectures, $[Ref^{RL}]$: real load data considered, DR: detection rate, DD: Detection delay, FPDR: False positive DR, DA: Detection accuracy, FPR: False positive rate, TPR: True positive rate, FDI: injected magnitude of FDI attack, payoffs: Game metric of attacker-defender cost in payoffs, SR: FDI attack sparsity ratio, SNR: Signal-to-noise ratio, MAPE: Mean absolute percentage error, PE: Percentage error between true and estimated states, AR: attacking rate (Attackability, or successful attacking probabilities), MSE: Mean square error. \end{tablenotes}}
\end{table*}

\begin{table*}
\begin{adjustbox}
{totalheight={20cm},keepaspectratio,gstore totalheight=\mylength}
\begin{tabular}{|m{0.1cm}|m{0.1cm}|m{2cm}|m{1.5cm}|m{0.10cm}m{0.10cm} m{0.10cm} m{0.10cm}m{0.10cm}|m{0.1cm}m{0.1cm}m{0.1cm}m{0.1cm} m{0.7cm}m{0.1cm}m{0.1cm}|m{2.65cm}|m{1cm}m{0.1cm}m{0.1cm}|}
 \hline
  & & & & \multicolumn{5}{c|}{\textbf{Attack model}}& \multicolumn{7}{c|}{\textbf{Attack target}} & & \multicolumn{3}{c|}{\textbf{Exp. platform}} \\      
 {\rotatebox{90}{\textbf{Category}}} & {\rotatebox{90}{\textbf{Subcategory}}} & \textbf{Algorithm} & \textbf{Reference} & {\rotatebox{90}{Complete information}}  & {\rotatebox{90}{Partial information}} & {\rotatebox{90}{LR attack}} & {\rotatebox{90}{GT attack}} & {\rotatebox{90}{Data-driven}} & {\rotatebox{90}{EMS}}   & {\rotatebox{90}{AGC}} & {\rotatebox{90}{DEM}} & {\rotatebox{90}{MMS}} & {\rotatebox{90}{Network comm.}} & {\rotatebox{90}{Intelligent device}} & {\rotatebox{90}{Renewable DER}} & \textbf{Perf. metric} & {\rotatebox{90}{Bus system}}  &
   {\rotatebox{90}{Simulation}} & {\rotatebox{90}{Test bed}} \\   \hline
{\multirow{35}{*} {\rotatebox{90}{\textbf{Protection-based defence}}}} & {\multirow{16}{*} {\rotatebox{90}{Optimal measurement selection}}}  & Greedy algorithm & \cite{7234893}$_{SR}^{D,c}$ & &$\checkmark$& & & & $\checkmark$ & &  & & SCADA & $\checkmark$& & DR vs SR, DR vs FAR & 9, 14, 57 & $\checkmark$ &   \\  \cline{4-20}
&  & & \cite{bobba2010Protection1}$_{SR}^{D,c}$ &$\checkmark$ && & & & $\checkmark$ & &  & & SCADA & $\checkmark$& & DR vs SR, protected sensors & multiple & $\checkmark$ &   \\  \cline{4-20}
& & &\cite{7370787}$^{D,d}$ & $\checkmark$&& & & & $\checkmark$ & &  & & SCADA & $\checkmark$& & \# of protected meters vs attack cost & multiple & $\checkmark$ & \\  \cline{4-20}
& & &\cite{5622046}$_{SR}^{D,c}$ & &$\checkmark$& & & & $\checkmark$ & $\checkmark$ &  & & SCADA &$\checkmark$ & & Attack cost vs \# of protected IEDs &14, 118 & $\checkmark$ & \\  \cline{4-20}
& & &\cite{6162362Protection2}$^{D,c}$ & $\checkmark$&& & & & $\checkmark$ &  &  & & SCADA &$\checkmark$ & & Optimal meter protections &14 & $\checkmark$ &   \\  \cline{3-20}
& & Game-theoretic & \cite{8219710}$^{D,c}$ & &&$\checkmark$ & & & $\checkmark$ &  &  & $\checkmark$& SCADA & $\checkmark$ & &Optimal meter protections &14, 30& $\checkmark$ & \\\cline{4-20}
& & & \cite{6378414}$^{D,c}$ & &&$\checkmark$ & & & $\checkmark$ &  &  & $\checkmark$& SCADA & $\checkmark$ & & Load shedding cost &5, 9, 14& $\checkmark$ &  \\  \cline{4-20} 
&& & \cite{7570183}$^{D,c}$ & &&$\checkmark$ & & & $\checkmark$ &  &  & $\checkmark$& SCADA &  & & Load shedding cost&9, 14& $\checkmark$ &   \\  \cline{4-20}
&& & \cite{esmalifalak2013bad}$^{D,c}$ & &&$\checkmark$ & & & $\checkmark$ &  &  & $\checkmark$& SCADA & $\checkmark$ & & AR, DR, LMP&5& $\checkmark$ &   \\  \cline{2-20}
& {\multirow{20}{*} {\rotatebox{90}{Grid topology perturbation}}} & MTD & \cite{6149267}$^{A,c}$ & $\checkmark$&& & & & $\checkmark$ &  &  & & SCADA &  & & Power loss &multiple& $\checkmark$ &   \\  \cline{4-20}
& & & \cite{6687991}$^{D,c}$ & $\checkmark$&& & & & $\checkmark$ &  &  & & SCADA &  & & DR, FDI &14& $\checkmark$ &  \\  \cline{4-20} 
& & & \cite{9144465}$^{D,c}$ & $\checkmark$&& & & & $\checkmark$ &  &  & & SCADA &  & & OPF cost, DR vs FAR &14& $\checkmark$ &  \\  \cline{4-20} 
& & & \cite{rahman2014moving}$^{D,c}$ & &$\checkmark$& & & & $\checkmark$ &  &  & & SCADA & $\checkmark$ & & AR &14& $\checkmark$ &  \\ \cline{4-20}
&& & \cite{8385215}$^{A,c}$ &$\checkmark$ && & & & $\checkmark$ &  &  & & SCADA &  & & DR vs attacked states  &6, 57& $\checkmark$ &   \\ \cline{4-20} 
& & & \cite{8820088}$^{A,c}$ & &$\checkmark$& & & & $\checkmark$ &  &  & & SCADA & & &Meter protection cost, PE vs FDI  &6, 14, 57& $\checkmark$ & \\  \cline{4-20} 
& & & \cite{8732433}$^c$ &$\checkmark$ && & & & $\checkmark$ &  &  & & SCADA &  & &DR vs FDI, TPR vs FAR  &39& $\checkmark$ &  \\  \cline{3-20}
& & Hidden MTD & \cite{8252724}$^{A,c}$ &$\checkmark$ && & & & $\checkmark$ &  &  & & SCADA & & &DR vs FDI, TPR vs FAR  &14& $\checkmark$ &  \\  \cline{4-20}
&   & & \cite{8762178}$^{D,c}$ & &$\checkmark$& & & & $\checkmark$ &  &  & & SCADA &  & &DR vs SR, DR vs Perturbation ratio  &multiple& $\checkmark$ &   \\  \cline{4-20} 
& & & \cite{zhang2020hiddenness}$^{D,c,RL}$ & $\checkmark$&$\checkmark$& & & & $\checkmark$ &  &  & & SCADA & & &DR vs SR, DR vs Perturbation ratio  &57, 118& $\checkmark$ &  \\  \cline{4-20}
&   & & \cite{8586470}$^{A,d}$ & $\checkmark$&$\checkmark$& & & & $\checkmark$ &  &  & & SCADA & $\checkmark$ &$\checkmark$ &Reactance rate, Power loss&66& $\checkmark$ &   \\  \hline {\multirow{18}{*} {\rotatebox{90}{\textbf{Statistical  model}}}} &
{\multirow{1}{*} {\rotatebox{90}{GLR}}}  & $\ell_1$-norm min. & \cite{6032057Centralised2}$^{D,c}$ &  $\checkmark$  &  & $\checkmark$ &  &  & $\checkmark$ & $\checkmark$ &  & $\checkmark$  &  &  &  & DR, FAR & 14 & $\checkmark$ & \\  \cline{3-20} 
& & Auto-regressive   
  & \cite{tang2016detection}$^{D,c}$  &  &  &  &  & $\checkmark$  & $\checkmark$ &  &  &   &  &  &  & DR, FAR & 30 & $\checkmark$ & \\ \cline{2-20}&
{\multirow{1}{*} {\rotatebox{90}{Bayesian}}}  & Game-theoretic & \cite{mangalwedekar2017bayesian}$^{A,c}$ &  &  &  &  & $\checkmark$  & $\checkmark$ &  &  &   &  &  &  & payoffs & 14 & $\checkmark$ &  \\  \cline{3-20}
& &Joint est. det.  
  & \cite{8101499}  &  $\checkmark$  &  &  &  &  &  &  &  & & WSN &  &  & DR, FAR & - & $\checkmark$ & \\  \cline{4-20}   
& & &\cite{7127816} &  $\checkmark$  &  &  &  &  & $\checkmark$  &  &  & & &  &  & MSE, FPR & - & $\checkmark$ & \\  \cline{4-20}   
& & & \cite{7086893}$^{D,c}$ &  $\checkmark$  &  &  &  &  & $\checkmark$  &  &  & & &  &  & MSE, FPR & - & $\checkmark$ &  \\ \cline{2-20} &
{\multirow{6}{*} {\rotatebox{90}{QCD}}}  & CUSUM & \cite{8278264}$^{D,cd}$  &  $\checkmark$  &  &  &  &  & $\checkmark$ &  & $\checkmark$ &   &  &  &  & DD, FDI & 14 & $\checkmark$ &   \\  \cline{4-20}   
& & & \cite{5766111}$^{A,c}$  &  $\checkmark$  &  &  &  &  & $\checkmark$ &  &  &   &  &  &  & DR, DD, FAR & 4 & $\checkmark$ &   \\  \cline{4-20} 
& & & \cite{6949126}$^{D,c}$ &  $\checkmark$  &  &  &  &  & $\checkmark$ & & & & & & & DD, FAR & multiple & $\checkmark$ &  \\  \cline{3-20}
& & Seq. change  
  & \cite{8808884}$^{D,c}$  &  $\checkmark$  &  &  &  &  & $\checkmark$ &  &  &   &  & $\checkmark$ &  & DD, FAR & 13 & $\checkmark$ & \\  \cline{4-20}   
& & &\cite{6982207}$^{D,d}$  & & $\checkmark$ & &  &  &  &  & $\checkmark$ &   & SCADA & &  & DR, meters and DD, FPR & 14 & $\checkmark$ &  \\  \cline{4-20}   
& & &\cite{8485421}$^{D,d}$  & & & &  & $\checkmark$ &  &  & $\checkmark$ &   & SCADA & &  &DD, FAR & 14 & $\checkmark$ & \\ \cline{2-20} &
{\multirow{4}{*} {\rotatebox{90}{Stat. dist.}}}  & KL dist.  & \cite{7035067}$^{A,c,RL}$  & $\checkmark$ & & &  & & $\checkmark$ &  &  &  & SCADA & &  &DR, FDI & 14 & $\checkmark$ &   \\  \cline{4-20}   
& & &\cite{7961272}$^{A,c,RL}$  & $\checkmark$ & & &  & & $\checkmark$ &  &  & &SCADA & &  &DR, FDI & 14 & $\checkmark$ & \\  \cline{3-20} 
& & JS dist. & \cite{8688077}$^{A,c,RL}$  & $\checkmark$ & & &  & & $\checkmark$ &  &  & & SCADA & &  &DR, FDI & 14 & $\checkmark$ & \\  \cline{4-20}   
& & &\cite{8600016}$^{A,c,RL}$  & $\checkmark$ & & &  & & $\checkmark$ &  &  & & AMI & &  &DR, FDI & 14 & $\checkmark$ & \\ \hline
\end{tabular} \end{adjustbox}
\end{table*} 

\begin{table*}
\begin{adjustbox}
{totalheight={19.5cm},keepaspectratio,gstore totalheight=\mylength}
\begin{tabular}{|m{0.15cm}|m{0.15cm}|m{2cm}|m{1.5cm}|m{0.10cm}m{0.10cm} m{0.10cm} m{0.10cm}m{0.10cm}|m{0.1cm}m{0.1cm}m{0.1cm}m{1.25cm} m{1cm}m{0.1cm}m{0.1cm}|m{2.75cm}|m{1cm}m{0.1cm}m{0.1cm}|} \hline
   & & & & \multicolumn{5}{c|}{\textbf{Attack model}}& \multicolumn{7}{c|}{\textbf{Attack target}} & & \multicolumn{3}{c|}{\textbf{Exp. platform}} \\      
 {\rotatebox{90}{\textbf{Category}}} & {\rotatebox{90}{\textbf{Subcategory}}} & \textbf{Algorithm} & \textbf{Reference} & {\rotatebox{90}{Complete information}}  & {\rotatebox{90}{Partial information}} & {\rotatebox{90}{LR attack}} & {\rotatebox{90}{GT attack}} & {\rotatebox{90}{Data-driven}} & {\rotatebox{90}{EMS}}   & {\rotatebox{90}{AGC}} & {\rotatebox{90}{DEM}} & {\rotatebox{90}{MMS}} & {\rotatebox{90}{Network comm.}} & {\rotatebox{90}{Intelligent device}} & {\rotatebox{90}{Renewable DER}} & \textbf{Perf. metric} & {\rotatebox{90}{Bus system}}  &
   {\rotatebox{90}{Simulation}} & {\rotatebox{90}{Test bed}} \\  \hline {\multirow{3}{*} {\rotatebox{90}{\textbf{Stat. model}}}} 
&{\multirow{1}{*} {\rotatebox{90}{Sparse recovery}}}  & Sparse matrix optimization & \cite{liu2013detection}$_{SR}^{D,c,RL}$  && & &&$\checkmark$ & $\checkmark$ &  &  & & SCADA & &  & TPR, FPR, SR, SNR & 57\& 118 & $\checkmark$ &   \\  \cline{4-20}   
& & &\cite{6740901SparseOpt1}$_{SR}^{D,c,RL}$ & & & & &$\checkmark$ & $\checkmark$ &  &  & & SCADA & &  & TPR, FPR, SR, SNR & 57\& 118 & $\checkmark$ &   \\  \cline{3-20} 
& &Fast Go Decomposition  & \cite{8489956}$^{D,c}$ & &$\checkmark$ & & &$\checkmark$ & $\checkmark$ &  &  & & PMU & & & DA, TPR, FPR & 118 & $\checkmark$ & \\ \hline 
 {\multirow{24}{*} {\rotatebox{90}{\textbf{Data-driven}}}} &
{\multirow{8}{*} {\rotatebox{90}{Supervised ML}}}  & SVM & \cite{7063894}$^{D,d}$ & & & & &$\checkmark$ &  &  &  & & PMU & & &  DA, P, R& 9,57,118 & $\checkmark$ &  \\  \cline{4-20}
& & &\cite{esmalifalak2014detecting}$^{A,c}$ & & & & &$\checkmark$ & $\checkmark$ &  &  & & SCADA & & & P, R, F1  & 118 & $\checkmark$ &  \\  \cline{4-20}
& &&\cite{8221908}$^{D,c}$ & & $\checkmark$& & &$\checkmark$ & $\checkmark$ &  &  & & SCADA & & &  F1, DR, FAR,& 118 & $\checkmark$ & \\  \cline{4-20}
& & &\cite{8555556} & & & & & & $\checkmark$ &  &  & &  & & & P, R, F1  & - & $\checkmark$ & $\checkmark$ \\  \cline{3-20}
& &ANN   
  & \cite{8221908}$^{D,c}$ & & $\checkmark$& & &$\checkmark$ & $\checkmark$ &  &  & & SCADA & & &  F1, DR, FAR,& 118 & $\checkmark$ & \\  \cline{4-20}   
& & &\cite{ganjkhani2019novel}$^{D,c}$ & & $\checkmark$ & $\checkmark$& & & & $\checkmark$ &  & $\checkmark$& SCADA & & &  P, MSE& 14 & $\checkmark$ & \\ \cline{4-20}
& & &\cite{8658084}$^{A,c,RL}$ &$\checkmark$ & && &&$\checkmark$  &  &  & & SCADA & & &AUC& 14 & $\checkmark$ & \\  \cline{3-20}
& &KNN  & \cite{7727361}$_{SR}^{D,c}$ &$\checkmark$ & && &&$\checkmark$  &  &  & & SCADA & & &F1, DA& 30 & $\checkmark$ &  \\ \cline{2-20}
&{\multirow{1}{*}{\rotatebox{90}{Semi-sup. ML}}}  & Semi-supervised ANN & \cite{7063894}$^{D,d}$ & & & & &$\checkmark$ &  &  & $\checkmark$ & & PMU & & &  DA, P, R& 9,57,118 & $\checkmark$ &  \\  \cline{3-20}   
& &Semi-supervised GMM  
  & \cite{8221908}$^{D,c}$ & & $\checkmark$& & &$\checkmark$ & $\checkmark$ &  &  & & SCADA & & &  F1, DR, FAR,& 118 & $\checkmark$ & \\  \cline{2-20}
&{\multirow{12}{*} {\rotatebox{90}{Deep learning}}}  & DFFNN & \cite{8450487}$^{A,c}$ & & $\checkmark$ & & & & $\checkmark$ & &  & & SCADA & & &  DA, P, R, TPR vs FPR&14 & $\checkmark$ &   \\ \cline{3-20}
& &CDBN & \cite{7926429}$_{SR}^{D,c,RL}$ & & & $\checkmark$ & & $\checkmark$&  & $\checkmark$ &  &$\checkmark$ & SCADA & & &  DA, TPR vs FPR&118,300 & $\checkmark$ &  \\  \cline{3-20}
& &DRNN   
& \cite{8334607}$^{A,c}$ & & $\checkmark$& & & & $\checkmark$ &  &  & & PMU & & &  DA, TPR, FPR&118,300 & $\checkmark$ & \\  \cline{4-20}
& & & \cite{dehghani2020deep}$_{SR}^{A,d}$ & & $\checkmark$& & & &  &  &  & & PMU & &$\checkmark$ &DA, TPR, FPR&118 & $\checkmark$ & \\  \cline{3-20} 
& &CNN   
  & \cite{8791598}$_{SR}^{D,c}$ & &$\checkmark$& & & &  $\checkmark$ &  &  & & SCADA & & &Location\& attack DA, TPR, FPR&14, 118 & $\checkmark$ & \\  \cline{4-20}
& & & \cite{9049087}$_{SR}^{D,c}$ & &$\checkmark$& & & &  $\checkmark$ &  &  & & SCADA & & &DA&39 & $\checkmark$ & \\  \cline{3-20}
  & &GAN   
  & \cite{9144530}$^{A,d}$ & &$\checkmark$& & & &  &  &  & & PMU & &$\checkmark$ &DA, P, R&13, 123 & $\checkmark$ &\\  \cline{2-20}
&{\multirow{4}{*} {\rotatebox{90}{RL}}}  & Q-learning & \cite{8248780}$^{A,cd,RL}$ & &$\checkmark$& & & & $\checkmark$ &  & $\checkmark$ & & SCADA &$\checkmark$ & &Voltage sag, DA&39 & $\checkmark$ & $\checkmark$ \\  \cline{3-20}  
& &SARSA   
  & \cite{8514804}$^{D,c}$ & && & $\checkmark$ & & $\checkmark$ &  &  & & - &$\checkmark$ & & DD, FAR, P, R&14 & $\checkmark$ &  \\  \cline{3-20}
& &Bayesian Bandit   
  & \cite{8735819}$^{D,c}$ & $\checkmark$&$\checkmark$& & & & $\checkmark$ &  &  & & SCADA & & &DR, MSE&14 & $\checkmark$ &  \\  \cline{2-20}
&{\rotatebox{90}{DRL}}  & Deep-Q-network & \cite{8786811}$^{A,c}$ & && & $\checkmark$ & &  & $\checkmark$ &  & $\checkmark$& - &$\checkmark$ & & DD, FAR&9,14,30 & $\checkmark$ &  \\  \hline 

{\multirow{20}{*} {\rotatebox{90}{\textbf{Prevention}}}} &
{\multirow{16}{*} {\rotatebox{90}{Cryptographic schemes}}}  & {\multirow{6}{*} {\rotatebox{90}{Encryption/Dec.}}} &\cite{5622046}$_{SR}^{D,c}$ & &$\checkmark$& & & & $\checkmark$ & $\checkmark$ &  & & SCADA &$\checkmark$ & & Attack cost vs \# of protected IEDs &14, 118 & $\checkmark$ & \\  \cline{4-20}
& & &\cite{6303199}$^{d}$ & && & & & & & $\checkmark$ & & Modbus, AMI &$\checkmark$ & $\checkmark$& Packet loss, Computational cost &- & $\checkmark$ & \\  \cline{4-20}
& & &\cite{7571177}$^{d}$ & && & & & & & $\checkmark$ & & DNP3 &$\checkmark$ & $\checkmark$& Packet loss, Communication overhead, Computational cost & - & $\checkmark$ &  \\  \cline{3-20}

& & {\multirow{8}{*} {\rotatebox{90}{Authentication}}}  
& \cite{hittini2020fdipp}$^{d}$ & && & & &$\checkmark$ & & $\checkmark$ & &SCADA, IEC 61850 &$\checkmark$ & &Delay, Packet loss, Comm. overhead, Comp. cost & - & $\checkmark$ &  \\  \cline{4-20} 
& &  &\cite{mahmood2016lightweight}$^{d}$ & && & & &$\checkmark$ & & $\checkmark$ & &AMI & & &Latency, Comm. overhead, Comp. cost & - & $\checkmark$ & \\ \cline{4-20}
& &  &\cite{8854144}$^{d}$ & && & & &$\checkmark$ & & $\checkmark$ & &AMI & & &Comm. overhead, Energy overhead, Comp. cost & - & $\checkmark$ & \\ \cline{3-20}
&&{\rotatebox{90}{Signature}}  & \cite{7822933}$^{c}$ & && & & &$\checkmark$ & &  & &Wireless comm., C37.118 &$\checkmark$ & &Signature overhead & 42 & $\checkmark$ & $\checkmark$ \\  \hline
\end{tabular} \end{adjustbox}
\end{table*}

\begin{table}
\begin{adjustbox}
{totalheight={8cm},keepaspectratio,gstore totalheight=\mylength}
\begin{tabular}{|m{0.13cm}|m{0.13cm}|m{0.13cm}|m{1.5cm}|m{0.10cm}m{0.10cm} m{0.10cm} m{0.10cm}m{0.10cm}|m{0.1cm}m{0.1cm}m{0.1cm}m{1.25cm} m{0.9cm}m{0.1cm}m{0.1cm}|m{2.7cm}|m{0.4cm}m{0.1cm}m{0.1cm}|} \hline
  & & & & \multicolumn{5}{c|}{\textbf{Attack model}}& \multicolumn{7}{c|}{\textbf{Attack target}} & & \multicolumn{3}{c|}{\textbf{Exp. platform}} \\      
 {\rotatebox{90}{\textbf{Category}}} & {\rotatebox{90}{\textbf{Subcategory}}} &
 {\rotatebox{90}{\textbf{Algorithm}}}
 & \textbf{Reference} & {\rotatebox{90}{Complete information}}  & {\rotatebox{90}{Partial information}} & {\rotatebox{90}{LR attack}} & {\rotatebox{90}{GT attack}} & {\rotatebox{90}{Data-driven}} & {\rotatebox{90}{EMS}}   & {\rotatebox{90}{AGC}} & {\rotatebox{90}{DEM}} & {\rotatebox{90}{MMS}} & {\rotatebox{90}{Network comm.}} & {\rotatebox{90}{Intelligent device}} & {\rotatebox{90}{Renewable DER}} & \textbf{Performance metric} & {\rotatebox{90}{Bus system}}  &
   {\rotatebox{90}{Simulation}} & {\rotatebox{90}{Test bed}}  \\  \hline  {\multirow{12}{*} {\rotatebox{90}{\textbf{Prevention}}}}
 &{\multirow{8}{*} {\rotatebox{90}{Blockchain-based defence}}}  & {\multirow{1}{*} {\rotatebox{90}{Data prot.}}} & \cite{8326530}$^{d}$ &$\checkmark$ && & & & & & $\checkmark$ &$\checkmark$ &AMI & & &AR vs manipulated meters & 118 & $\checkmark$ &   \\  \cline{4-20}  
& &  &\cite{mbarek2020enhanced}$^{d}$ &$\checkmark$ && & & & & & $\checkmark$ &$\checkmark$ &AMI & &$\checkmark$ &Transaction verification of energy supply-demand& - & $\checkmark$ &  \\  \cline{3-20}
&  &{\rotatebox{90}{Privacy preserv.}}   
& \cite{8918446}$^{c}$ & && & & &$\checkmark$ & & & &SCADA & &&DA, DR vs FAR& - & $\checkmark$ &  \\  \cline{4-20}  
& &  &\cite{chen2020blockchain}$^{d}$ & && & & & & &$\checkmark$ & &Smart meter & &&Transaction delay, Computational cost& - & $\checkmark$ & \\ \hline
\end{tabular}
\end{adjustbox}
\end{table}
\subsection{Evaluation Criteria}
In order to quantify the efficacy and associated challenges of the different defence strategies, several key evaluation criteria against the proposed algorithms are suggested in relation to the requirements of the power systems and the Smart Grid cybersecurity. The assessment criteria used to compare the selected defence algorithms against the FDI attacks are summarized in Table \ref{criteriaTable}. 

One of the main evaluation criteria is defence algorithm, a criterion that reflects the reviewed defence techniques. Five commonly used attack construction methodologies have been considered for the attack model criterion, namely attack with complete information, attack with partial information, load redistribution (LR) attack, grid topology (GT) attack, and attack using data-driven approaches have been considered. Furthermore, the AC and DC models are considered for the power flow model. The reviewed articles are also evaluated from network-centric point of view (considering centralised and decentralised architecture). Additionally, the FDI attack defence papers are investigated with regards to the numerous cyber-physical entities of the Smart Grid. Consequently,  seven major Smart Grid components, including EMS, AGC, DEM, MMS, network \& communications, intelligent devices, and renewable resources, were identified  for attack target evaluation criterion. Notice that the different components of the Smart Grid can be seen from the discussion in Section \ref{backG}. Finally, two evaluation criteria, namely performance metrics and experimental platform have been inspected. The evaluation criteria are used to compare and contrast among the various defence strategies as detailed in Section \ref{comparisonSec} and summarised in Table \ref{cmpTable}. 
\section{Comparison and Statistics Among Defence Strategies}\label{comparisonSec}
In our review paper, 111 publications are considered for the defence class. Here, the various countermeasure strategies are compared and some statistical facts based on the evaluation criteria are presented. 
\subsection{The Defence Strategies}
The conventional BDDs, namely $\chi^2$, LNR, and detection based on SE partitioning are merely used for bad data processing (see Section \ref{bdds} for detail). Consequently, the literature considered in this class did not take into account FDI attacks. Nevertheless, they have been blended with a variety of other approaches for detecting the FDI attacks and serve as the basis for most countermeasure techniques. For example, the $\chi^2-$detector and the LNR detector have been employed in the detection based on dynamic SE subcategory.    

Data-driven and detection based on statistical models are the two most popular defence categories comprising just under half of the total, with the former standing at approximately one-quarter and the latter at 23\% of the total. Because of the complexity of Smart Grid infrastructure, the sheer volume of data, and the fact that high-performance computing devices are becoming available, data-driven techniques are increasingly powering various applications of the smart power system. As a result, plenty of data-driven defence techniques especially DL and RL have been pursued these days as means of developing more effective detection against the FDI attacks (as demonstrated in Section \ref{dataDriven}). The optimal placement of PMU, optimal selection of measurement quantities, MTD, all under the protection-based category are the other prominent defence strategies against the false data attacks in Smart Grid cybersecurity (standing at 21\%). Prevention-based defences (cryptographic functions and Blockchain technologies) are among the emerging security control mechanisms against the incumbent cyberattacks. Especially, these techniques are popular across the demand side management (i.e. consumption-side) of the Smart Grid.       
\subsection{Performance Metric}
The defence strategies vary, among other things, in terms of algorithmic design, adversarial method, attack target, and network architecture. For this reason, instead of providing a distinct performance metrics for all the attack countermeasures, we present comprehensive qualitative metrics. A plentiful of performance metrics are presented for each of the countermeasure subcategories (see $17^{th}$ column of Table \ref{cmpTable}). For example, across the protection-based defence category, optimal subset of meter, optimal IED protection, and attack cost are the main metrics considered. Further, packet loss, computational cost, communication cost, and end-to-end delay are the main evaluation metrics adopted among the prevention schemes. In most of the detection based on dynamic SE, statistical-based models, and data-driven defence categories, detection rates (in terms of probability of detection, True Positive Rate (TPR)) are compared against False Positive Rates (FPR) or False Alarm Rates (FAR).  
\subsection{Experimental Platform}
The vast majority of studies performed numerical results based on simulations of IEEE standard or modified electric grid test cases. Various sizes of test cases have been considered, IEEE 14 bus system being the most widely referred test case. Although the vast majority of literature use only a single test case to conform their numerical results, some considered multiple test cases. To further verify the efficacy of their proposal some scholars used a real-time load data, most of which used a dataset from the New York Independent System Operator. Almost all of the studies are based on simulations using MATPOWER\footnote{https://matpower.org/}, a MATLAB power system toolbox, and few others utilise PowerFactory\footnote{https://www.digsilent.de/en/powerfactory.html} and TOMLAB\footnote{https://tomopt.com/tomlab/} optimization toolbox. Finally, very a few incorporated co-simulations and hardware testbeds.
\section{Main Gaps of Existing Defence Strategies Against the FDI Attacks} \label{litGap}
While detailed research reviews of each defence category have been addressed in Section \ref{countermeasure}, in what follows, we describe the key gaps of existing defence researches.

\textbf{Some Emerging Smart Grid Areas Are Not Well Studied}:
The plethora of literature examined in this review paper tried to cover a multitude of Smart Grid infrastructures; however, there are some open issues with respect to the scope (network architecture, DERs, and communication systems). The majority of existing countermeasure researches have focused on the traditional centralised EMS. While decentralized energy generation and distribution systems (such as the DERs) have become very popular, they have been among the most vulnerable cyber-physical components to FDI attacks. But, only a few research studies have been undertaken with respect to defence strategies of the DERs. This can be seen from the $16^{th}$ column of Table \ref{cmpTable}. Further, only few papers have discussed in the SAS, AMI, and WAMS-based  communication systems.

Moreover, it has been described (see section \ref{prevDef}) that preventive security measures are essential in the fight against FDI attacks in the Smart Grid. Especially, lightweight cryptography and blockchain-based security systems are the least studied areas.

Throughout this report it has been mentioned that the power system measurement data can reveal anomaly in the face of cyberattacks. It is also highly likely that physical faults contribute to the abnormal functioning of the power system. Therefore, the research on FDI attack can be extended with respect to the identification between the cyber attacks and power physical faults. Differentiating between the cyber threats and the physical faults can be beneficial for the operators as it helps them to react against unnecessary losses. Only very few researches \cite{7063234} \cite{anwar2015data} are done in this respect. Especially, a real-time detection scheme is required considering the sparsity of FDI attack and low-dimensional
 property of the measurement data received at the control center.
 
\textbf{General Shortcomings of the Countermeasures: Performance, Computational Cost, and Feasibility of Deployment}: 
The conventional BDD-based detection methods have not been able to handle stealthy and sparse FDI attacks and are thus vulnerable to the FDI attack. Therefore, the numerous defence algorithms analysed in the literature have achieved much stronger security controls against the incumbent cyberattack. There are, however, certain limitations that are worth mentioning here. For example, in spite of their potential to defend key grid components against the bad data injection attacks, the protection-based defence schemes have certain drawbacks: First, deployment of PMUs in the large-scale Smart Grid is much more expensive, especially with the emerging ubiquitous sensing infrastructure. It has also been shown \cite{shepard2012evaluation} that PMUs are susceptible to the injection of false data attacks via GPS spoofing, which requires a more appealing security scheme. Additionally, determining the subset of measurements is a large-complexity problem. MTD can allow grid operators to proactively protect measurements from malicious attackers by introducing perturbations to network data or topology that can inevitably lead to uncertainties and costs against adversaries. Yet, the MTD protection approaches can be compromised by intelligent FDI attackers if the attackers can identify MTD changes before they perform the attack (as has been demonstrated in  \cite{8252724} \cite{8762178}). 

It has been seen that detection based on dynamic state estimators are more powerful countermeasure techniques than the BDDs; however, the use of WLS and Kalman filter-based signal processors incurs an immense computational burden. Physics-aware data-driven defence approaches, on the other hand, are much more robust for power system security, especially for dynamically changing power system variables. Besides that, the prevalence of GPUs makes it practical to satisfy the computational requirements of advanced ML models. Currently, deep neural networks, reinforcement learning, and the convergence of the two are the most favoured ML models for detecting the FDI cyberattacks.

Although missing currently, using commercial-level datasets of stealthy FDI attacks can practically verify the efficacy of data-driven countermeasure techniques.

\textbf{Need for Corroboration of Experimental Results Via Testbed Platform}:
Although the literature surveyed in this paper have proven their cybersecurity solutions via numerical simulations benchmarked against standardised test cases, it is vital to validate the experimental results via cyber-physical testbeds, which is missing in the literature except to a few of them (\cite{8555556} \cite{8248780} \cite{7822933}).
This downside can be seen from the perspectives of data- and system-oriented approaches. Most of the FDI attack schemes surveyed did not consider commercial-level datasets, which otherwise, can practically validate the vulnerability of the state estimators to the stealthy FDI attacks. Even more significant, the countermeasure techniques can also incorporate real-world off datasets.

Testbeds \cite{6473865} are essential tools for testing the performance evaluation of algorithms and protocols in the Smart Grid. The highly complex and multidisciplinary essence of the Smart Grid requires the implementation of cyber-physical testbeds with different characteristics for comprehensive experimental validation. There is a considerable need to analyse new Smart Grid security concepts, architectures, and vulnerabilities via cyber-physical system test platforms. Recently, there is a growing attention in the study of cyber-physical Smart Grid testbeds \cite{6473865} \cite{7740849}. Most notably,  hardware-in-the-loop \cite{7177085} test platforms have become much more popular for the development, analysis, and testing of cyber-physical components of the electrical power system. For example, some Smart Grid stakeholders, such as ABB\footnote{https://new.abb.com/news/detail/62430/abbs-acs6000-power-electronics-grid-simulator-pegs-tests-medium-voltage-equipment}, Siemens Power Technologies\footnote{https://assets.new.siemens.com/siemens/assets/api/uuid:1fb8264a-9ee6-4d71-a703-bb68beb7ca94/version:1587982708/rtds-datasheet-en-1909.pdf}, and OPAL RT\footnote{https://www.opal-rt.com/hardware-in-the-loop/} foster hardware-in-the-loop testing using real-time digital simulators across various Smart Grid realms, including microgrids, SAS- and WAMS-based protection environments. Therefore, we suggest that assessing the effects of FDI attacks on the Smart Grid using the hardware-in-the-loop testbed platform is critical in crafting the stringent cybersecurity requirements. 
\section{Emerging Advanced Applications: Future Research Directions} \label{futurePros}
Securing the electricity grid is one of the highest priorities of many countries around the world. Academic studies and industries are expected to tackle a range of issues for future research on cyber defence in the Smart Grid infrastructure. Particularly, the reliance of reliable and secure power system operation on the communication infrastructure, along with potential cyber threats are increasingly growing. In the following, potential emerging advanced applications are discussed as means of future research prospects.

\textbf{Cybersecurity for Emerging Smart Grid Communication Systems}: 
Despite the fact that the communication infrastructure is the most critical target to the FDI attacks, the countermeasures have to be studied well, especially, across the SAS-compliant IEC 61850 and the WAMS-compliant IEEE C37.118. The FDI attack can well be studied especially with the incorporation of cyber-physical testbed platforms \cite{6473865}. Moreover, although AMI is one of the most vulnerable communication systems to the FDI attack, little has been done on the defence against this attack. 
Especially, given the increasing adoption of IoT in the Smart Grid, it will be interesting to address cybersecurity issues of IoT-based AMI with regard to the FDI attacks. Software-defined networking \cite{diro2018differential} is one of the emerging networking applications. The coupling of software-defined networking with the Smart Grid applications can bring efficient network monitoring. However, the security issue of this technology is worth investigating especially with respect to the FDI attacks. Further, it has been indicated \cite{8735819} that cognitive radio can help the implementation of a control-sensing mechanism to identify and account for the detection of the FDI attacks in the Smart Grid. In addition, countermeasures against FDI attacks on heterogeneous cognitive radio, WSN and IoT are potential cybersecurity researches which are worth investigating. Specifically, the application of data-driven models along with the countermeasure strategies across the more intelligent communication arena of the Smart Grid seems to be a promising solution in tackling against the orchestrated cyberattacks.

\textbf{Security Framework Based on Lightweight ML}: Countless memory and computational-restricted wireless sensor nodes are connected to IoT applications in Smart Grid. Several reports have shown that such limitations raise obstacles to the usage of conventional security measures over IoT systems. Security frameworks using lightweight ML \cite{sliwa2020limits} can be proposed for resource-constrained IoT devices. For example, lightweight ML can be proposed for prevention schemes such as encryption, message authentication, and dynamic key management against the false data attacks in an end-to-end Smart Grid communication system.   

\textbf{FDI Attack Detection in Edge Computing}: The growing popularity of distributed renewable energy generation requires reduced processing costs and communication overheads. In a distributed computing environment, edge computing \cite{diro2018distributed} improves the communication overhead and system bandwidth by bringing the processing and data storage near to the origin of data source. Further, the emergence of Industry 4.0 across a number of industries, including the Smart Grid, brings ubiquitous networked elements, and intelligent edge computing. Although the intelligent edge computing is expected to be able to meet the needs of the ever-growing IoT users in the Smart Grid, there are inherent security threats. For example, bringing more of such IoT devices to the edge network can introduce various cybersecurity threats. FDI attacks can be challenging in edge computing environment. Distributed detection using DL or deep RL against the incumbent attacks can be a potential research direction in edge computing-based Smart Grid. 

\textbf{Blockchain Technology}: A Blockchain-based defence for privacy preservation and anomaly detection in Smart Grid is a very new research area, which requires a further investigation.
\section{Conclusion}\label{cncln}
Smart Grid faces a growing threat from an emerging cyber-physical attack called FDI. By injecting a stealthy falsified attack vectors, adversaries can infringe critical Smart Grid information, may render the power system unobservable, and may culminate in large-scale failure of the power system operation. 

This survey paper analysed extensive review of existing state-of-the-art researches on cyber defence against the incumbent cyberattacks in Smart Grid. A taxonomy of five major categories and subcategories of the different countermeasures was proposed. Furthermore, in order to quantify the efficacy and associated challenges of the various proposed algorithms in the literature surveyed, a number of key evaluation criteria was used in relation to the requirements of the power systems and the Smart Grid cybersecurity. Finally, future research directions for mitigation techniques of the FDI attacks were also proposed as a way of advancing the Smart Grid cybersecurity framework.
\bibliographystyle{srt}
\bibliography{Haftu_main.bbl} 
\end{document}